\documentclass[runningheads]{llncs}
\usepackage[utf8]{inputenc}
\usepackage[T1]{fontenc}
\usepackage[english]{babel}
\usepackage{amssymb,amsmath}
\usepackage[font=small,center]{caption}
\usepackage{xcolor}
\usepackage{graphicx}
\usepackage{wrapfig}
\usepackage{makecell}
\usepackage{proof}
\usepackage{wasysym}
\usepackage{fontawesome}
\usepackage{tabularx}
\usepackage{xspace}
\usepackage{multirow}
\usepackage{tikz}
\usetikzlibrary{positioning,arrows,shapes,hobby,backgrounds,calc,trees,fit}
\usetikzlibrary{arrows,automata,shadows,shapes.arrows}
\usepackage{tikz-cd}
\pgfdeclarelayer{background}
\pgfsetlayers{background,main}
\usepackage{indentfirst}
\usepackage{arydshln}
\usepackage{makebox}
\usepackage{varwidth}
\usepackage{proof}
\usepackage{mathtools}
\usepackage{stackengine}
\usepackage{bussproofs}
\usepackage{float}

  {\gdef\scalefactor{#1}\begin{center}\proofSkipAmount \leavevmode}%
  {\scalebox{\scalefactor}{\DisplayProof}\proofSkipAmount \end{center} }

\makeatletter
\RequirePackage[bookmarks,unicode,colorlinks=true]{hyperref}%
   \def\@citecolor{blue}%
   \def\@urlcolor{blue}%
   \def\@linkcolor{blue}%

\def\orcidID#1{\smash{\href{http://orcid.org/#1}{\protect\raisebox{-1.25pt}{\protect\includegraphics{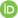}}}}}
\makeatother

\begin{document}
%
%
\title{Symbolic Path-guided Test Cases \\
for Models with Data and Time} 
%
\titlerunning{Symbolic Path-guided Test Cases for Models with Data and Time}
\author{
Boutheina Bannour \inst{1}
\and
Arnault Lapitre \inst{1}
\and
Pascale Le Gall\inst{2}
\and
\\
Thang Nguyen\inst{2} 
}

\authorrunning{ B. Bannour, A. Lapitre, P. Le Gall, and T. Nguyen}
\institute{
Université Paris-Saclay, CEA, List, F-91120, Palaiseau, France
\and
Université Paris-Saclay, CentraleSupélec, MICS, F-91192, Gif-sur-Yvette, France
}
\maketitle

\begin{abstract}
This paper focuses on generating test cases from timed symbolic transition systems. At the heart of the generation process are symbolic execution techniques on data and time. Test cases look like finite symbolic trees with verdicts on their leaves and are based on a user-specified finite symbolic path playing the role of a test purpose. Generated test cases handle data  involved in time constraints and uninitialized parameters, leveraging the advantages of symbolic execution techniques.
\end{abstract}

\newcommand{\includeMacros}{ 42 }

\newcommand\tc[1][]{\text{TC}}

\newcommand\tclk[1][]{clock}

\newcommand{\IFresh}[1]{IFresh{#1}}
\newcommand{\OFresh}[1]{OFresh{#1}}
\newcommand{\OdFresh}[1]{OdFresh{#1}}
\newcommand{\OCond}[1]{OCond{#1}}

\newcommand{\commentbb}[1]{\textcolor {red}    {{ BB@:  {#1}}}}
\newcommand{\commentplg}[1]{\textcolor{orange} {{ PLG@: {#1}}}}
\newcommand{\commental}[1]{\textcolor {olive}  {{ {\scriptsize @AL:}  {#1}}}}
\newcommand{\commenttn}[1]{\textcolor {magenta}{{    TN@:  {#1}}}}

\newcommand{\maxWaitingBound}{
\text{TM}
\xspace}

\definecolor{falured}{rgb}{0.5, 0.09, 0.09}
\newcommand{\f}[1]{\textcolor{black}{\bf #1}}

\newcommand{\vartime} { 
tb 
}

\newcommand{\ATMID} { 
\text{ATM\_ID}
}
\newcommand{\ACCEPT}{ 
\text{ACCEPT}
}

\newcommand{\chan}[1]{ \text{#1} }

\newcommand{\actIn}[1] { 
\text{#1} ?
}
\newcommand{\actOut}[1]{ 
\text{#1} ! 
}

\newcommand{\reset}[1]{\{ {#1} \}}

\newcommand{\guard}[1]{ [ {#1} ] }

\newcommand{\subst}[1]{ \langle {#1} \rangle}


\newcommand{\trans}[4]{$ 
\xrightarrow{ {#1}, \guard{#2}, \reset{#3}, \subst{#4} }
$}

\newcommand{\freshOut}[3]{ y^{#1}_{{#2}_{#3}} }

\newcommand{\freshOutSingle}[1]{ y_{#1} }

\newcommand{\piAlias}[1]{ \textcolor{black}{ \pi( ec_{#1} ) } }

\newcommand{\obs  } { 
{obs}
}
\newcommand{\spec  } { 
{spec}
}

\newcommand{\uncIn } { 
{ucIn}
}
\newcommand{\unspec} { 
{uspec}
}

\newcommand{\builtTransition}[6]{
    \begin{center}
    \setlength\extrarowheight{6pt}
    \small
    
    \begin{tabular}{|c|c|c|c|c|}
    \hline
        \textbf{Source} &
        \textbf{Clocks} &
        \textbf{Action} &
        \textbf{Substitution} &
        \textbf{Target}
    \\ \hline
        {#1} & 
        {#2} & 
        {#4} & 
        {#5} & 
        {#6}   
    \\ \hline
        \multicolumn{5}{ |c| }{\multirow{1}{*}{ \shortstack[c]{ \textbf{Guard} } }}
    \\ \hline
        \multicolumn{5}{ |c| }{\multirow{1}{*}{ \shortstack[c]{ {#3}} }}
    \\ \hline
    \end{tabular}
    
    \end{center}
}

\newcommand{\builtTransitionAlt}[8]{
    \begin{center}
    \setlength\extrarowheight{6pt}
    \small
    
    \begin{tabular}{r|c|c|c|c|c|}
    \cline{2-6}
        & \textbf{Source} & \textbf{Clocks} & \textbf{Action} & \textbf{Substitution} & \textbf{Target}
    \\ \hline
        \multicolumn{1}{ |r| }{ if $post(ec) \neq \epsilon$ } &
        \multirow{2}{*}{\shortstack[c] {#1}} & 
        {#2} & 
        \multirow{2}{*}{\shortstack[c] {#5}} & 
        \multirow{2}{*}{\shortstack[c] {#6}} & 
        {#7}   
    \\ \cline{1-1} \cline{3-3} \cline{6-6}
        \multicolumn{1}{ |r| }{ else } &
               & 
        {#3} & 
             & 
             & 
        {#8}   
    \\ \hline
        & \multicolumn{5}{ c| }{\multirow{1}{*}{ \shortstack[c]{ \textbf{Guard} } }}
    \\ \cline{2-6}
        & \multicolumn{5}{ c| }{\multirow{1}{*}{ \shortstack[c]{ {#4}} }}
    \\ \cline{2-6}
    \end{tabular}
    
    \end{center}
}

\definecolor{BLUEColor}{RGB}{171, 216, 231}

\tikzstyle{State}=[circle, minimum width=0.9cm, inner sep=0pt, outer sep=0pt]
\tikzstyle{NotBelongTP_EC}=[circle,fill,color=BLUEColor,text=black]
\tikzstyle{Pass_EC}=[circle, fill, color=green, text=black]
\tikzstyle{Inconc_EC}=[circle, fill, color=orange!80, text=black]
\tikzstyle{Fail_EC}=[circle, fill, color=white, text=black]
\tikzstyle{Middle_EC}=[circle, minimum width=0.9cm, inner sep=0pt, outer sep=0pt, ball color=yellow]
\tikzstyle{DataQuiescence_EC}=[circle, minimum width=0.9cm, inner sep=0pt, outer sep=0pt, ball color=orange]
\tikzstyle{TimedQuiescence_EC}=[circle, minimum width=0.9cm, inner sep=0pt, outer sep=0pt, ball color=green]
\tikzstyle{Skip_EC}=[circle, minimum width=0.9cm, inner sep=0pt, outer sep=0pt, ball color=yellow]
\tikzstyle{line}=[draw, -latex, line width=1pt]
\section{Introduction}

{\em Context.} Symbolic execution~\cite{King76,GastonGRT06,FrantzenTW06,BoerB19} explores programs or models' behaviors using formal parameters instead of concrete values and computes a logical constraint on them, the so-called path condition.  Interpretations of these parameters satisfying the constraint yield inputs that trigger executions along the desired path. 
Symbolic execution's primary application is test case generation, where considering test cases guided by different symbolic paths facilitates achieving high coverage across diverse behaviors. Symbolic execution has been defined first for programs~\cite{King76} and extended later to models~\cite{GastonGRT06,FrantzenTW06,AndradeMJM11,StypBS10,BannourEGG12,TretmansTap19} in particular to symbolic transition systems where formal parameters abstract values of uninitialized data variables~\cite{GastonGRT06,BannourEGG12}, values of received data from the system's environment~\cite{GastonGRT06,FrantzenTW06,StypBS10,AndradeMJM11,BannourEGG12,TretmansTap19,AichernigT16}, and durations stored in clock variables~\cite{StypBS10,BannourEGG12}. 

{\em Contribution.} In this paper, we investigate test case generation from models given as symbolic transition systems that incorporate both data and time. Time is modeled with explicit clock variables, which are treated as a particular case of data variables that occur in guards and constrain the transitions' firing. Our approach allows for general logical reasoning that mixes data and time through symbolic execution, typically compared to Timed Automata~\cite{AlurD1994}, which are models dedicated to time and use tailored zone-based abstraction techniques to handle time. Test cases are built based on a test purpose, defined as a selected symbolic path of the model. 
We require that test purposes are deterministic in that any system behavior given as a trace cannot be executed both on the test purpose and on another symbolic path. By leveraging this determinism property and symbolic execution techniques, we define test cases as tree-like structures~\cite{Jeron09,Marsso0S18}, presenting the following advantages: (i) data and time benefit from comparable property languages, seamlessly handled with the same symbolic execution techniques; (ii) input communication channels are partitioned into 
controllable input channels enabling the test case to stimulate the system under test, and 
uncontrollable input channels enabling observation of data from third parties; (iii) state variables do not need to be initialized, and finally (iv), test cases can be easily executed on systems under test, typically achieved through behavioral composition techniques, such as employing TTCN-3~\cite{TTCN}, or by maintaining a test case state at runtime using on-the-fly test case execution~\cite{FrantzenTW04,GastonGRT06,TretmansTap19}. In either case, our test cases are to be coupled with constraint solving to assess the satisfiability of test cases' progress or verdict conditions. We provide a soundness result of our test case execution on the system under test in the framework of the timed conformance relation tioco~\cite{KrichenT04} issued from the well-established relation ioco~\cite{iocoTretmans96}. Finally, we implement our test case generation in the symbolic execution platform Diversity~\cite{diversity}.

{\em Paper plan.} We devote Sec.\ref{sec:Timed Input/Output Symbolic Transition Systems} to present timed symbolic transition systems mixing data and time, and in Sec.\ref{sec:Symbolic execution of TIOSTS}, we define their symbolic execution serving as the foundation for our test case generation. In Sec.\ref{sec:Conformance}, we give the main elements of the testing framework: the conformance relation tioco and test purposes. In Sec.\ref{sec:testcase}, we detail the construction of symbolic path-guided test cases. In Sec.\ref{sec:RelatedWork}, we provide some links to related work. In Sec.\ref{sec:Conclusion}, we provide concluding words.

\vspace{-.25cm}
\section{Timed Input/Output Symbolic Transition Systems}
\label{sec:Timed Input/Output Symbolic Transition Systems}


{\em Preliminaries on data types.}
For two sets $A$ and $B$, we denote $B^A$,  the set of applications from $A$ to $B$. We denote $\coprod_{i\in\{1,\ldots n\}} A_i$ the  disjoint union of sets  $A_1$, \ldots, and $A_n$. 
For a set $A$, $A^*$ (resp. $A^+$) denotes the set of all (resp. non-empty) finite  sequences of elements of $A$, with $\varepsilon$ being  the empty sequence. For any two sequences $w,w'\!\in\! A^*$, we denote $w.w'\!\in\! A^*$  their concatenation.

A data signature is a pair $\Omega=(S,Op)$ where $S$ is a set of type names and $Op$ is a set of operation names provided with a profile in $S^+$. 
We denote $V=\coprod_{s\in S}V_s$ the set of typed variables in $S$
 with  $\mathit{type}:V\rightarrow S$ the function that associates  variables with their type. 
The set $\mathcal{T}_{\Omega}(V)=\coprod_{s\in S}\mathcal{T}_{\Omega}(V)_{s}$ of $\Omega$-terms in $V$ is inductively defined over  $V$ and operations $Op$ of $\Omega$ as usual and the function $type$ is extended to $\mathcal{T}_{\Omega}(V)$ as usual.
 The set $\mathcal{F}_{\Omega}(V)$ of typed equational $\Omega$-formulas over $V$ is inductively defined over  the classical equality and inequality predicates $t \Join t'$ with $\Join\, \in \{ < , \leq , = , \ge , > \}$ 
 for any $t,t'\!\in\! \mathcal{T}_{\Omega}(V)_{s}$ and over usual Boolean constants and connectives $True$, $False$, $\neg$, $\vee$, $\land$ and quantifiers $\forall x$, $\exists x$ with $x$ a variable of $V$. We may use the syntax $\exists \{ x_1, \ldots, x_n \}$ for the expression $\exists x_1 \ldots \exists x_n$. 
 A substitution over $V$ is a type-preserving application $\rho: V \rightarrow \mathcal{T}_{\Omega}(V)$. 
The identity substitution over $V$ is denoted $id_V$ and substitutions are canonically extended on terms and formulas. 
 
 An $\Omega$-model $M = (\coprod_{s\in S}M_s, (f_M)_{f \in Op})$ provides a set of values $M_s$ for each type $s$ in $S$ and a concrete operation  $f_M:M_{s_1}\times\dots\times M_{s_n}\rightarrow M_s$ for each operation name $f : s_1 \ldots s_n \rightarrow s$ in $Op$. 
  An interpretation $\nu: V \rightarrow M$ associates a value in $M$ with each variable $v\!\in\! V$, 
and is canonically extended  to $\mathcal{T}_{\Omega}(V)$ and $\mathcal{F}_{\Omega}(V)$ as usual. For $\nu$ an interpretation in $M^V$, $x$ a variable in $V$ and $v$ a value in $M$, $\nu[x \mapsto v]$ is the interpretation $\nu' \in M^V$ which sends $x$ on the value $v$ and coincides with $\nu$ for all other variables in $V$. 
For $\nu\in M^V$ and $\varphi\in \mathcal{F}_{\Omega}(V)$, the satisfaction of $\varphi$ by $\nu$ is  denoted  $M\models_{\nu}\varphi$ and is inductively  defined w.r.t. the structure of $\varphi$ as usual. We say a formula $\varphi\!\in\! \mathcal{F}_{\Omega}(V)$ is  satisfiable, denoted $Sat(\varphi)$, if there exists $\nu\!\in\!M^V$ such that $M\models_{\nu}\varphi$. 

In the sequel, we consider  a data signature $\Omega=(S,Op)$ with $time \in S$ to represent durations and $Op$ containing the usual operations  $<:time.time\rightarrow Bool$ and $+:time. time\rightarrow time$,~\ldots An $\Omega$-model $M$ being  given, $M_{time}$ is denoted $D$ and is isomorphic to the set of non-negative real numbers. $<:time.time\rightarrow Bool$ and $+:time.time\rightarrow time$ are mapped to their usual meanings. 

\noindent
{\em Timed Input/Output Symbolic Transition Systems (TIOSTS)} are automata handling data and time, and defined over a {\em signature} $\Sigma = (A,K,C)$, 
where: 
\begin{itemize}
    \item $A= \coprod_{s\in S}A_s$ and $K$ are pairwise disjoint sets of variables representing respectively \emph{data variables} and \emph{clock variables} of type $\mathit{time}$ 
    \item and $C= \coprod_{s\in S}C_s$ is a set of \emph{communication channels} with the convention $type(c) = s$ for any $c \in C_s$. 
    Moreover, channels of type $s\!\in\! S$  are partitioned into input and output channels, i.e., $C_s\!=\! C_s^{in}\coprod C_s^{out}$. 
\end{itemize}
We denote $C^{in}\!=\!\coprod_{s\in S} C_s^{in}$, resp. $C^{out}\!=\!\coprod_{s\in S} C_s^{out}$, the set  of all input, resp. output, channels, regardless of their type.


Interactions of TIOSTS  with the  environment are expressed in terms of communication actions.
The \emph{set of communication actions} over $\Sigma$  is $Act(\Sigma)=I(\Sigma)\cup O(\Sigma)$ where:
\begin{itemize} 
\item $I(\Sigma)=\{c?x~|~c\!\in\! C^{in}, x\!\in\! A_{type(c)}\}$ is the set of input actions, and
\item $O(\Sigma)=\{c!t~|~c\!\in\! C^{out}, t\!\in\! {\cal T}_\Omega(A \cup K)_{type(c)}\}$ is the set of output actions.
\end{itemize}


$c?x$ denotes the reception of a value to be stored in $x$ through channel $c$. $c!t$ denotes the emission of the value corresponding to the current interpretation of term $t$ through channel $c$. 
The \emph{set of concrete communication actions} over $C$  is $Act(C)=I(C)\cup O(C)$ where: \\
\centerline{$I(C)=\{c?v~|~c\!\in\! C^{in},v\!\in\! M_{type(c)}\}$  and $O(C)=\{c!v~|~c\!\in\! C^{out},v\!\in M_{type(c)}\}$ }
{\em Notations.} For $a\in Act(\Sigma)$ (resp. $a\in Act(C)$) of the form $c?y$ or $c!y$, $chan(a)$ and $val(a)$ denote $c$ and $y$ respectively. For expressiveness concerns,  we also use extensions of those 
actions: either carrying $n$ pieces of data, i.e.  $c!(t_1,\ldots,t_n)$ or $c?(x_1,\ldots,x_n$), and simple signals $c!$ or $c?$ which are actions carrying no-data. 

\begin{definition}[TIOSTS]
\label{def:tiosts}
A  TIOSTS over  $\Sigma=(A,K,C)$ is a triple $\mathbb{G}\!=\!(Q, q_0, \mathit{Tr})$, where 
\begin{itemize}
    \item $Q$ is the set of states, 
    \item $q_0\!\in\! Q$ is the initial state, \item $\mathit{Tr}$ is the set of transitions of the form $(q,act,\phi,\mathbb{K},\rho,q')$ with $q, q'\in Q$,  $act\in Act(\Sigma)$, $\phi \in \mathcal{F}_{\Omega}(A\cup K)$, $\mathbb{K}\subseteq K$ and $\rho: A\rightarrow {\cal T}_\Omega(A\cup K)$ is a type-preserving function. 
\end{itemize}    
\end{definition}
In the sequel, given a transition $tr$ of the form $(q,act,\phi,\mathbb{K},\rho,q')$, we will access its components by their name, for example, $act(tr)$ for its communication action. We comment on the ingredients of a TIOSTS through the TIOSTS given in Ex.~\ref{ex:TIOSTS}.

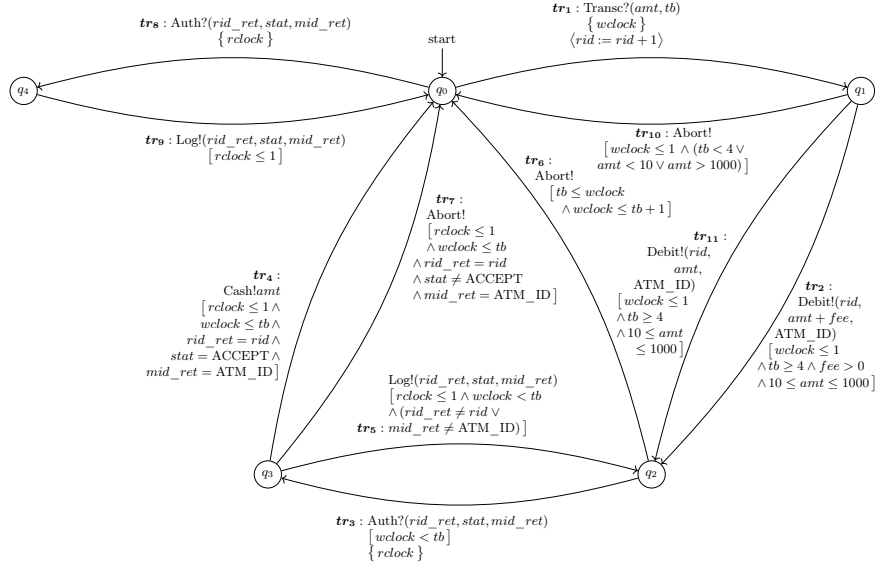
\begin{figure}[t]
     \centering
\vspace{-.2cm}
     
     \resizebox{.975\textwidth}{!}{
     \hspace{-.7cm}
%
\ifx\includeMacros\undefined

    \definecolor{falured}{rgb}{0.5, 0.09, 0.09}
    \newcommand{\f}[1]{\textcolor{falured}{\bf #1}}

    \newcommand{\actIn}[1] { \textcolor{orange}{\text{#1} ?} }
    \newcommand{\actOut}[1]{ \textcolor{violet}{\text{#1} !} }

\else

    \renewcommand{\f}[1]{ {\bf #1} }

    \renewcommand{\actIn}[1] { \text{#1} ? }
    \renewcommand{\actOut}[1]{ \text{#1} ! }
\fi
%
%
\newcommand{\freset}[1]{ \big\{\, {#1} \,\big\} }

\newcommand{\fguard}[1]{ \big[\, {#1} \,\big] }

\newcommand{\bguard}{ \big[\, }
\newcommand{\eguard}{ \,\big] }

\newcommand{\fsubst}[1]{ \big\langle {#1} \big\rangle }
%

    \begin{tikzpicture}
    [->,
    xscale=1.675,yscale=1.675,
    semithick,font=\small]
    \tikzstyle{every state}=[minimum size=5mm,fill=white,draw
    ]


    \node[state, initial above] (a_0) at (-0.5,1.5) {$q_0$};
    \node[state] (a_1) at (5.5,1.5) {$q_1$};

    \node[state] (a_2) at (2.5,-4) {$q_2$};
    \node[state] (a_3) at (-3,-4) {$q_3$};




    \path (a_0) edge [bend left=15,above]   (a_1);

    \path (a_1)
    edge [bend left=15,right,near start]
    (a_2);

    \path (a_1)
    edge [bend right=15,right,near start]
    (a_2);

    \path (a_1)
    edge [bend left=15,right,near start]
    (a_0);

    \path (a_2) edge [bend right=15]   (a_0);

    \path (a_2) edge [bend left=15,right,near start]  (a_3);

    \path (a_3)
    edge [bend left=15, left,near end]
    (a_0);

    \path (a_3) edge [bend right=15,left,near start]
    node [below right = -17mm and -3mm] {
    $
    $
    } (a_0);

    \path (a_3) edge [bend left=15,left,near end]  node [below left = -8mm and -11mm] {
    $
    $
    } (a_2);

    \node[state] (a_4) at (-6.5,1.5) {$q_4$};
    \draw  (a_0) edge[bend right =15]  (a_4);
    \draw  (a_4) edge[bend right =15] (a_0);


    \node at (2.0,2.45) {
        $\begin{array}{c}
            \boldsymbol{tr_1}: \actIn{Transc}(amt, \vartime)
        \\
            \freset{ wclock }
        \\
            \fsubst{ rid:=rid+1}
        \end{array}$
    };

    \node at (4.875,-2) {
        $\begin{array}{l}
            \qquad\quad\;\; \boldsymbol{tr_2}:
        \\
            \qquad\quad \actOut{Debit} (rid,
        \\
            \qquad\; amt+fee,
        \\
            \quad\; \ATMID)
        \\
            \;\;\, \bguard wclock \leq 1
        \\
            \land\, \vartime \geq 4
            \land fee > 0
        \\
            \land\, 10 \leq amt \leq 1000
            \eguard
        \end{array}$
    };

    \node at (2.7184,-1.4055) {
        $\begin{array}{l}
            \qquad\quad \qquad\;\;\, \;\;\, \boldsymbol{tr_{11}}:
        \\
            \qquad\quad \actOut{Debit} (rid,
        \\
            \qquad\quad\qquad amt,
        \\
            \quad\quad \ATMID)
        \\
            \;\;\, \bguard wclock \leq 1
        \\
            \;\;\,  \land\,  \vartime \geq 4
        \\
            \;\;\,  \land\, 10 \leq amt
            \\
            \;\;\, \;\;\,  \;\;\,\leq 1000
            \eguard
        \end{array}$
    };

    \node at (-0.5,-4.9062) {
        $\begin{array}{cl}
            \boldsymbol{tr_3}: & \actIn{Auth}(rid\_ret, stat, mid\_ret)
        \\
            & \fguard{ wclock < \vartime }
        \\
            & \freset{ rclock }
        \end{array}$
    };

    \node at (-3.7812,-1.5) {
        $\begin{array}{r}
        \\ \\ \\
            \boldsymbol{tr_4} :
        \\
            \actOut{Cash}amt
        \\
            \bguard rclock \leq 1 \,\land
        \\
            wclock \leq \vartime \,\land
        \\
            rid\_ret = rid \,\land
        \\
            stat=\ACCEPT \,\land
        \\
            mid\_ret = \ATMID
            \eguard
        \end{array}$
    };

    \node at (-0.2808,-2.9988) {
        $\begin{array}{cl}
            & \actOut{Log}(rid\_ret, stat, mid\_ret)
        \\
            & \bguard rclock \leq 1 \land wclock < \vartime
        \\
            & \land\, (rid\_ret \ne rid \;\lor
        \\
            \boldsymbol{tr_5}: & mid\_ret \ne \ATMID)
            \eguard
        \end{array}$
    };

    \node at (1.75,0.15) {
        $\begin{array}{l}
            \boldsymbol{tr_6}:
        \\
        \;\;\,   \actOut{Abort}
        \\
        \quad   \;\;\,  \bguard \vartime \le wclock \\
        \quad   \quad\;\;\, \land\, wclock \leq \vartime + 1
        \eguard
        \end{array}$
    };

    \node at (0.125,-0.7496) {
        $\begin{array}{l}
            \quad\;\;\,  \boldsymbol{tr_7}:
        \\
            \quad   \actOut{Abort}
        \\
            \quad  \bguard rclock \leq 1
        \\
            \quad   \land\, wclock \leq  \vartime
        \\
            \land\, rid\_ret = rid
        \\
            \land\, stat \neq \ACCEPT
        \\
            \land\, mid\_ret = \ATMID
            \eguard
        \end{array}$
    };

    \node at (-3.3184,2.35) {
        $\begin{array}{c}
            \boldsymbol{tr_8}: \actIn{Auth}(rid\_ret, stat, mid\_ret)
        \\
            \freset{ rclock }
        \end{array}$
    };

    \node at (-3.3184,0.675) {
        $\begin{array}{c}
            \boldsymbol{tr_9}: \actOut{Log}(rid\_ret, stat, mid\_ret)
        \\
            \fguard{ rclock \leq 1 }
        \end{array}$
    };

    \node at (2.8176,0.6559) {
        $\begin{array}{c}
            \boldsymbol{tr_{10}}:
            \actOut{Abort}
        \\
            \bguard 
            wclock \leq 1 \,\land
            (\vartime < 4 \,\lor
        \\
            amt < 10
            \lor amt > 1000)
            \eguard
        \end{array}$
    };

\end{tikzpicture}
      }
     \caption{Example TIOSTS of an ATM 
     }
     \label{fig:tiosts}
     \vspace{-.7cm}
 \end{figure}




\begin{example}\label{ex:TIOSTS} The TIOSTS $\mathbb{G}\!=\!(Q,q_0,Tr)$ in Fig.~\ref{fig:tiosts} represents a simple Automatic Teller Machine (ATM)
 with $Q\!=\!\{q_0,\dots,q_4\}$ and $Tr\!=\! \{tr_1,\ldots,tr_{11} \}$. Its signature introduces two clocks ($wclock$, $rclock$), 7 data variables ($rid$, $amt$, $\vartime$, $fee$, $rid\_ret$, $stat$, $mid\_ret$)  and 6 channels including 2 input channels (
 $\chan{Transc}$, $\chan{Auth}$) and 4 output channels ($\chan{Debit}$, $\chan{Abort}$, $\chan{Cash}$ and $\chan{Log}$).
 
 \noindent Transition $tr_1: q_0$\resizebox{!}{11pt}{ 
\trans{\actIn{Transc} {(amt, \vartime)}} {True} {wclock} {rid:=rid+1} } $q_1$ represents a reception  on channel \chan{Transc} of a client withdrawal request for a given amount  stored in variable $amt$ and the corresponding bound on processing time stored in variable $\vartime$, which can vary due to bank security checks. 
 $tr_1$ is unconditionally  fired (due to the guard $True$), resets clocks in $\mathbb{K}\!=\!\{wclock\}$ and updates variable $rid$ with ${rid}\!+\!1$. $tr_1$ abstracts client interaction and bank processing time retrieval.

\noindent Transition $tr_2 : q_1$\resizebox{!}{11pt}{ 
\trans{\actOut{Debit}(rid,amt+fee,\ATMID)} {wclock\leq 1 \land \vartime \geq 4 \land fee > 0 \land 10 \leq amt \leq 1000} {} {} } $q_2$
 represents an emission of (bank) debit request on channel Debit of the value of $rid$, the value of the term $amt+fee$, and the value of the constant $\ATMID$. $tr_2$ can be fired if and only if the duration since ${wclock}$ reset is less than or equal~$1$, the processing time bound is greater than or equal~$4$, the ATM $fee$ is strictly positive, and the withdrawal amount in some range (between $10$ and $1000$).

 Other transitions represent debit authorization reception ($tr_3$), cash return ($tr_4$), logging non-involved debit authorization ($tr_5$ and $tr_{9}$), cancellation upon timeout ($tr_6$) or debit refusal ($tr_7$), cancellation due to amount out of range or inappropriate processing time bound ($tr_{10}$), reception of non-involved debit authorization ($tr_8$), and feeless debit request ($tr_{11}$).
\end{example}

\vspace{-.25cm}
\section{Symbolic execution of TIOSTS}
\label{sec:Symbolic execution of TIOSTS}

We use symbolic execution 
techniques for defining the semantics of TIOSTS: transitions are executed not for concrete values but rather using fresh variables
and computing constraints on them. Given an TIOSTS $\mathbb{G} = (Q,q_0,Tr)$ over $\Sigma = (A,K,C)$,
we consider a set $F$ of fresh variables disjoint from TIOSTS variables, i.e. $F \cap (A\cup K)=\emptyset$, and
partitioned with the following subsets:
\begin{itemize}
\item $F^{ini}$ a set of variables dedicated to initialize variables of $\mathbb{G}$~;
\item $F^{in} = (F^{in}_c)_{c\in C^{in}}$  verifying that variables in $F^{in}_c$ are of type $type(c)$; 
\item $F^{out} = (F^{out}_c)_{c\in C^{out}}$  verifying that variables in $F^{out}_c$ are of type $type(c)$;
\item $F^{dur}$ a set of variables of type $time$.
\end{itemize}


For the signature $\Sigma_F\!=\!(F,\emptyset,C)$, the set $Evt(\Sigma_F)$ of {\em symbolic events} over $\Sigma_F$ is  $F_{time} \times (Act(\Sigma_F) \cup \{\_\})$  with  $\_$ for indicating the absence of an action. For $ev=(z,act)$ in $Evt(\Sigma_F)$, $delay(ev)$ and $act(ev)$ denote resp. $z$ and $act$. Intuitively, $z$ is the duration elapsed between the action preceding $act$ and $act$.  

 An Execution Context (EC) $ec$ is a data structure of the form $(q, \pi, \lambda, ev, pec)$ composed of pieces of information about symbolic execution:
\begin{itemize}
    \item $q \in Q$, a state (control point) of the TIOSTS reached so far, 
    \item $\pi \in {\cal F}_{\Omega}(F)$, a constraint on variables in $F$, the so-called \emph{path condition}, to be satisfiable by the symbolic execution to reach $ec$, 
    \item $\lambda : A \cup K \rightarrow {\cal T}_{\Omega}(F)$, a substitution, 
    \item $ev \in Evt(\Sigma_F)$, a symbolic event that has been executed to reach $ec$, 
    \item $pec$ a predecessor of $ec$ useful to build a symbolic tree in which nodes are ECs and edges connect predecessor ECs to ECs themselves. 
\end{itemize}
For any execution context $ec$, we note $q(ec), \pi(ec), \lambda(ec), ev(ec)$ and $pec(ec)$ to denote the corresponding elements in $ec$. In the same line, we also note $act(ec)$, $delay(ec)$ and $chan(ec)$ for resp. $act(ev(ec))$, $delay(ev(ec))$ and $chan(act(ev(ec)))$. For convenience, $Sat(\pi(ec))$ will be denoted $Sat(ec)$. 


\noindent Initial ECs are of the form $ec_0=(q_0,True,\lambda_0,\_,self)$ 
with: $\lambda_0$ associating to every variable of $A$ a distinct fresh variable of $F^{ini}$, and to variables of $K$ the constant $0$; "$\_$" an identifier indicating the absence of an action to start the system; and $self$ an identifier indicating that
, the predecessor of an initial EC is the initial context itself. $\mathbb{EC(G)}$ denotes the set of all ECs of a TIOSTS $\mathbb{G}$. 

\noindent For a non-initial execution context $ec$, we use the notation $pec(ec)$ \resizebox{!}{11pt}{$ \xrightarrow[]{ev(ec)}$} $ec$ or $pec(ec)$ \resizebox{!}{11pt}{$ \xrightarrow[]{ev(ec)}$} $ec \in \mathbb{EC}$ if $ec$ and its predecessor are both in a subset $\mathbb{EC}$ of $\mathbb{EC}(\mathbb{G})$. 

Transitions are executed symbolically from an EC. An example execution on the TIOSTS of Fig.~\ref{fig:tiosts} is provided before presenting the general definition.


\noindent \begin{minipage}[b]{.65\linewidth}
\begin{example} 
\label{ex:symb-ex}
In Fig.\ref{fig:symbex-succ-ec}, the symbolic execution of transition $tr_2$ (Ex.\ref{ex:TIOSTS}) is depicted from the execution context $ec_1$, where variables $fee$, $rid$, $rid\_ret$, $stat$, and $mid\_ret$ are evaluated with fresh initial parameters. This execution creates a successor context $ec_2$, where the clock $wclock$ is associated with a new duration $\f{z_1}$ in $F^{dur}$, representing the time passed since the preceding event. The emission event
 $(\f{z_1},\actOut{Debit}(\f{\freshOut{1}{D}{1}},\f{\freshOut{1}{D}{2}},\f{\freshOut{1}{D}{3}}))$ with $\f{\freshOut{1}{D}{1}},\f{\freshOut{1}{D}{2}},\f{\freshOut{1}{D}{3}}$ variables resp. in $F^{out}_{\chan{Debit},1},F^{out}_{\chan{Debit},2},F^{out}_{\chan{Debit},3}$
 corresponds to the symbolic evaluation of the action $act(tr_2)\!=\!\actOut{Debit}(rid,amt+fee,\ATMID)$.  
 
 The evaluation of the transition guard $\phi(tr_2)\!=\!wclock\!\le\!1\!\land\!\vartime\! \geq\!4 \!\land\!fee\!>\!0\!\land\!10\!\leq\!amt\!\leq\!1000$ yields formula $\f{z_1}\!\le\!1\!\land\!\f{\vartime_1}\!\geq\!4 \!\land\!\f{fee_0}\!>\!0\!\land\!10\!\leq\!\f{amt_1}\!\leq\!1000$ which together with identification conditions $\f{\freshOut{1}{D}{1}}\!=\!\f{rid_0} +1$, $\f{\freshOut{1}{D}{2}}\!=\!\f{amt_1}+\f{fee_0}$, $\f{\freshOut{1}{D}{3}}\!=\!\ATMID$ constitutes the path condition of $ec_2$. Identification conditions result from the transition action evaluation.
 \end{example}
\end{minipage}
\begin{minipage}[b]{0.45\linewidth}
\begin{figure}[H]
            \centering
    \resizebox{.7\textwidth}{!}{
\hspace{-1.2cm}
%





 \begin{tikzpicture}
    [xscale=1.2,yscale=1.2,
    ->,>=stealth',shorten >=1pt,auto,
    level/.style={sibling distance = 5.7cm,level distance = 1.8cm},
    ]
    \tikzstyle{every state}=[minimum size=5mm,fill=white,draw]
  
    	
    	

\node
at (-0.5,1) {
$
\begin{array}{|rl|}
\hline
    	q(ec_1): & q_1
    	\\
        \pi(ec_1): & True 
    	\\
    	\lambda(ec_1) : & rid \mapsto \f{rid_0} + 1, amt \mapsto \f{amt_1}, 
    	\\&
    	fee \mapsto \f{fee_0},     	 
    	rid\_ret \mapsto \f{rid\_ret_0}, 
    	\\ &
        stat \mapsto \f{stat_0},
        \\ &
        mid\_ret \mapsto \f{mid\_ret_0}, 
    	\\ &
    	\vartime \mapsto \f{tb_1},
    	wclock \mapsto 0,
        rclock \mapsto \f{z_0}
        \\
        ev(ec_1) : & (\f{z_0}, \actIn{Transc}( \f{amt_1} , \f{\vartime_1} ) )
    	\\
    	pec(ec_1) : &ec_0
    \\
     \hline
          \multicolumn{2}{|c|}{
               ec_1 \xrightarrow[]{(\f{z_1}, \actOut{Debit}(\f{\freshOut{1}{D}{1}},\f{\freshOut{1}{D}{2}},\f{\freshOut{1}{D}{3}}))} ec_2
          }
        \\
    \hline
    	q(ec_2): &q_2    
     \\
    \pi(ec_2) : & (\f{z_1} \le 1)
    \land (\f{\vartime_1} \geq 4)
    \\&
    \land\, (\f{fee_0} > 0)
    \\&
    \land\, (10 \leq \f{amt_1} \leq 1000)
    \\&
    \land\, (\f{\freshOut{1}{D}{1}}=\f{rid_0} +1) 
    \\&
    \land\,  
    ( \f{\freshOut{1}{D}{2}}=\f{amt_1}+\f{fee_0}) 
    \\&
    \land\, (\f{\freshOut{1}{D}{3}}=\text{ATM\_ID}) 
    \\  
    \lambda(ec_2) : &
    rid \mapsto \f{rid_0} + 1,
    amt \mapsto \f{amt_1},
     \\&
    fee \mapsto \f{fee_0},  
    rid\_ret \mapsto \f{rid\_ret_0}, 
    \\&
    stat \mapsto \f{stat_0}, \vartime \mapsto \f{\vartime_1}, 
    \\&
	mid\_ret \mapsto \f{mid\_ret_0}, 
 \\&
	wclock \mapsto \f{z_1},
	
    rclock \mapsto \f{z_0}+\f{z_1}
	\\
	ev(ec_2): & (\f{z_1}, \actOut{Debit} (\f{\freshOut{1}{D}{1}},\f{\freshOut{1}{D}{2}},\f{\freshOut{1}{D}{3}}))
    	\\
	pec(ec_2) : & ec_1
	\\
	\hline
    	\end{array}
    	$
};

\end{tikzpicture}
}
    \caption{Symbolic execution \\ of a TIOSTS transition }
    \label{fig:symbex-succ-ec}
    \vspace{-1cm}
\end{figure}
\vspace{.25cm}
\end{minipage}

Def.~\ref{def:tiosts-symbex} will make precise the EC successors' computation from TIOSTS transitions. While in Ex.~\ref{ex:symb-ex}, we have illustrated the symbolic execution of a unique transition ($tr_2$), we will define symbolic execution by simultaneously executing all the outgoing transitions from a given EC. 
In this way, we can introduce the same symbolic variables
for all outgoing transitions as far as they have the same role.
Generically, given an execution context $ec$ in $\mathbb{EC(G)}$, we will access the symbolic variables introduced by the executions from $ec$ with the following notations:
$f^{in}_c(ec)$ for $c\in C^{in}$, $f^{out}_c(ec)$ for $c\in C^{out}$, and $f^{dur}(ec)$. For convenience, all such fresh variables are available by default with every execution context $ec$, 
even if there is no outgoing transition from $q(ec)$ carrying on a given channel $c$. For $\alpha \in \{in,out\}$, $f^\alpha(ec) = \{ f_c^\alpha(ec) ~|~ c \in C^\alpha \}$. In Def~\ref{def:tiosts-symbex}, to make it easier to read, $f^{in}_c(ec)$, $f^{out}_c(ec)$ and $f^{dur}(ec)$ are respectively denoted as $x_c$, $y_c$ and $z$.


\begin{definition}[Symbolic Execution of a TIOSTS]\label{def:tiosts-symbex} 
Let $\mathbb{G}\!=\!(Q, q_0, \mathit{Tr})$ be a TIOSTS, $ec=(q,\pi,\lambda,ev,pec)$  be an execution context in  $\mathbb{EC(G)}$,  and $x_c$ be a fresh variable in $F^{in}_c$ for any $c\in C^{in}$,
$y_c$ be a fresh variable in $F^{out}_c$ for any $c\in C^{out}$ and let $z$ be a fresh  variable in $F^{dur}$. 

The successors of $ec$ are all execution contexts of the form $(q',\pi',\lambda',ev',ec)$ verifying that there exists a transition $tr=(q,act,\phi,\mathbb{K},\rho, q')$ in $Tr$ and constituents  $\lambda'$, $ev'$ and $\pi'$ of $ec'$ are defined as follows:

\begin{itemize}
\item the substitution $\lambda' : A \cup K \rightarrow {\cal T}_{\Omega}(F)$: 
\begin{equation}
\label{eq:sub_se}
\lambda'(w)  =
\begin{cases}
    \lambda'_0(\rho(w)) & \text{\bf if}\ w \in A  
    \\ 
    0 & \text{\bf if}\ w \in \mathbb{K}
    \\
    \lambda'_0(w) & \text{\bf else}
       \text{ i.e. } \text{if}\ w \in K \setminus \mathbb{K}
\end{cases}
\end{equation}

where $\lambda'_0 : A \cup K \rightarrow {\cal T}_{\Omega}(F)$ is the auxiliary function defined as:

\begin{equation}
\label{eq:subi_se}
\lambda'_0(w) =
\begin{cases}
    x_c & \text{\bf if}\ act = c?w  
    \\
    \lambda(w)+z &  \text{\bf if}\ w \in  K  
    \\
    \lambda(w) & \text{\bf else} 
    \end{cases}
  \end{equation}  
\item $ev'$ is $(z,c?x_c)$ if $act = c?w$ and is $(z,c!y_c)$ if $act = c!t$ for a given channel $c$

 
\item $\pi'$ is the formula
 $\pi \wedge \lambda'_0(\phi)$ if $act = c?w$ and is $\pi \wedge \lambda'_0(\phi) \wedge (y_c= \lambda'_0(t))$ if $act = c!t$ for a given channel $c$.  

\end{itemize}

The symbolic execution $SE(\mathbb{G})$ of $\mathbb{G}$ is  a couple $(ec_0, \mathbb{EC})$  where: $ec_0$ is an arbitrary initial EC, and $\mathbb{EC}$ is the smallest set of execution contexts containing $ec_0$ and all successors of its elements.
\end{definition}

The computation of the successors of an execution context translates the standard execution of a transition from that context: $\lambda'_0$ is an intermediate substitution that advances all clocks by the same fresh duration $z$ to indicate time passing, assigns to a data variable $w$ a fresh variable $x_c$ if $w$ is the variable of a reception ($c?w$) and leaves the other data variables unchanged. Then, $\lambda'$ is defined, for the data variables, by applying $\lambda'_0$  on the terms defined by the substitution $\rho$ introduced by $tr$ and, for the clock variables, by resetting the variables of $\mathbb{K}$ to zero and advancing the other clocks using $\lambda'_0$. 
The event action is either $c?x_c$ (case of a reception $c?w$) or $c!y_c$ (case of an emission $c!t$). The path condition $\pi'$ is obtained by the accumulation of the condition $\pi$ of the predecessor context $ec$ and of the guard $\phi$ of the transition  evaluated using $\lambda'_0$. 
Moreover, in case of an emission, $\pi'$ keeps track of the identification condition matching $y_c$ with the evaluation $\lambda'_0(t)$ of the emitted term $t$. 

In the sequel, we will denote $tr(ec)$ the transition that allows building the execution context $ec$. By convention, $tr(ec)$ is undefined for initial contexts.

\begin{figure}[t
]
     \centering

     \vspace{-.3cm}
     \resizebox{.95\textwidth}{!}{
      \hspace{-.7cm}\input{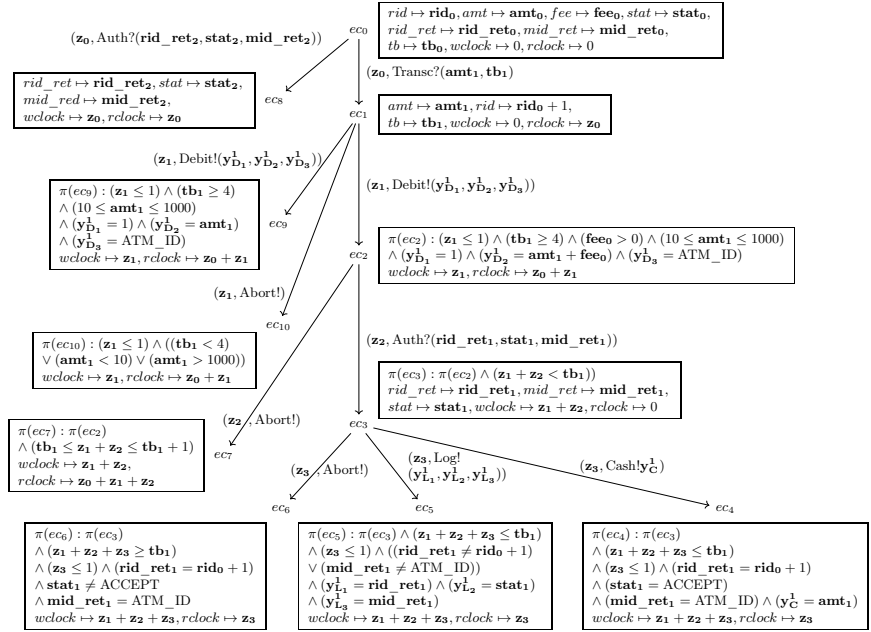}
      }
     \caption{Symbolic execution of the ATM TIOSTS of Fig.~\ref{fig:tiosts} 
     }
     \label{fig:tiosts-symbex}
      \vspace{-.7cm}
 \end{figure}

\begin{example}\label{ex:tiosts-symbex} Fig.~\ref{fig:tiosts-symbex} illustrates parts of the symbolic execution of the ATM example TIOSTS given in Fig.~\ref{fig:tiosts}.
\end{example}







So far, we have defined the symbolic execution of a TIOSTS without any adjustments related to our testing concerns from TIOSTS.
In the following, we will complete the symbolic execution with quiescent configurations, i.e., identifying situations where the system can remain silent. The system is usually expected to react by sending messages when it receives a message from its environment.
However, sometimes, it cannot emit an output from any given state \cite{iocoTretmans96,RusuMJ05,GastonGRT06,BannourEGG12,JanssenT19}. In such a case, the inability of the system to react
becomes a piece of information.  To make this fact clear 
 we enrich symbolic execution by adding a special output action $\delta!$ to denote the absence of output in those specific deadlock situations.





\begin{definition}[Quiescence enrichment]\label{def:Reactive_TIOSTS}
The quiescence enrichment $SE(\mathbb{G})^\delta$ of $SE(\mathbb{G})$ $= (ec_0, \mathbb{EC})$ is  $(ec_0, \mathbb{EC}^\delta)$ where $\mathbb{EC}^\delta$ is the set $\mathbb{EC}$ enriched by new execution contexts $ec^\delta$. For each context $ec=(q,\pi,\lambda,ev,pec)$ in $\mathbb{EC}$, a new context $ec^\delta=(q,\pi \land \pi^\delta,\lambda,(f^{dur}(ec),\delta!),ec)$ is considered where\footnote{with the convention that $\bigwedge$ quantified over empty conditions is the formula $True$}: 

\vspace{-.3cm}
\[\pi^\delta=\bigwedge_{\substack{
pec(ec')=ec\\ chan(ec')\in C^{out}} }\big(\forall f^{dur}(ec).\forall f^{out}_{chan(ec')}(ec) 
.\neg\pi(ec')\big) \]
\end{definition}

Let us emphasize that $\pi^\delta$ is satisfiable only for contexts $ec$ where there is no choice of values for the variables for triggering from $ec$ a transition carrying an emission. The context could not be considered quiescent if such a choice were possible, i.e. if there would exist an output transition towards an execution context $ec'$, for which there is an instantiation of variables $f^{dur}(ec)$ and $f^{out}_{chan(ec')}(ec)$ making the condition $\pi(ec')$ true.

\begin{example}\label{ex:quiescence} We discuss some examples from Fig.~\ref{fig:tiosts-symbex}. The execution context $ec_0$ does not have successors with outputs ($\pi^\delta$ is $True$) which denotes that the ATM is awaiting withdrawal requests or non-involved debit authorizations. This quiescent situation is captured by adding the context  $ec^\delta_0\!=\!(q_1,True,\lambda(ec_0),(\f{z_1},\delta!),ec_0)$. The execution context $ec_1$ has three successors with outputs, $ec_2$, $ec_9$ and $ec_{10}$. Then, $\pi^\delta$ is:
\vspace{-.3cm}
{\small
\[ \bigwedge_{j \in \{2, 9, 10 \}}
\forall f^{dur}(ec_1).\forall f^{out}_{chan(ec_j)}(ec_1).  \neg \pi(ec_{j}) 
\]
}
which is not satisfiable. Thus, there is no need to add a quiescent transition from $ec_1$. The same applies to $ec_2$ and $ec_3$. 

\end{example}

A symbolic path of $SE(\mathbb{G})^\delta=(ec_0, \mathbb{EC}^\delta)$ is a sequence $p = ec_0.ec_1\ldots ec_n$ where $ec_0$ is the initial context, for $i\in [1,n]$, $ec_i\in \mathbb{EC}^\delta$, and $pec(ec_i) = ec_{i-1}$. $Paths(SE(\mathbb{G})^\delta)$ denotes the set of all such paths. We will use the notation $tgt(p)$ to refer to $ec_n$, the last context of $p$.
We define the set of traces of a symbolic path $p$ in $Paths(SE(\mathbb{G})^\delta)$ by:
\vspace{-.275cm}
\[
    Traces(p) = \bigcup_{\nu\in M^{ F } } 
    \left\{
        \; \nu(p) ~|~ M \models_\nu \exists F^{ini}.\pi(tgt(p)) \; 
    \right\}
\]
where $\nu$ applies to a path $p$ of the form $p'.ec$ as $\nu(p)=\nu(p').\nu(ev(ec))$  with the convention $\nu(ev(ec_0))=\epsilon$ and 
$\nu(ev(ec)) = (\nu(z), c ? \nu(x))$ (resp. $(\nu(z), c ! \nu(y))$ or $(\nu(z), \delta!)$) 
if $ev(ec)$ is of the form  $(z,c?x)$ (resp. $(z,c!y)$ or $(z,\delta!)$).

By solving the path condition of a given path, we can evaluate all symbolic events occurring in the path and extract the corresponding trace. 
The set of traces of $\mathbb{G}$ is defined as :
\vspace{-.275cm}
\[
    Traces(\mathbb{G}) = \bigcup_{ p \in Paths(SE(\mathbb{G})^\delta )} Traces(p)
\]
\vspace{-.75cm}

\vspace{-.25cm}
\section{Conformance testing}
\label{sec:Conformance}
Conformance testing aims to check that a system under test behaves correctly w.r.t a reference model, a TIOSTS $\mathbb{G}$ in our case. The test case stimulates the system with inputs and observes the system's outputs, their temporalities, and the quiescent situations to compare them to those specified by $\mathbb{G}$. For generality, we propose test cases that control some inputs of the system under test while leaving other inputs driven by systems in its environment. 
We assume that the test case selects inputs and observes outputs on some channels while it can only observe inputs and outputs on other channels of the system under test.
Illustrating with the ATM, a test case provides the ATM with withdrawal requests
, and observes withdrawal authorizations received from the bank.

We characterize 
a \emph{Localized System Under Test (LUT)} (terminology from~\cite{GastonHG13,BGHLG17}) tested in a context where some inputs are not controllable. For this, we partition $C^{in} = CC^{in} \amalg UC^{in}$ where:
\begin{itemize}
    \item $CC^{in}$ is the set of controllable input channels and 
    \item $UC^{in}$ is the set of uncontrollable input channels. 
\end{itemize}
For a set of channels $C$, we denote $Evt(C)=D \times Act(C)$ the set of all concrete events that can occur in a trace: 
an event $(d,act)$ indicates that the occurrence of action $act$ happens $d$ units of time after the previous event. In 
model-based testing, a LUT is a black box and as such, is abstracted by a set of traces $LUT \subseteq Evt(C_\delta)^*$ with 
$C_\delta = C \cup \{\delta\}$ satisfying additional hypotheses, denoted as $\cal{H}$, ensuring its consistency.  Hypotheses ${\cal H}$ gather the following 3 properties, where for $\sigma_1, \sigma_2$  in $Evt(C)^*$, $d$ in  $D$ and $ev \in Evt(C)$ we have:

    $-$ {\em stable by prefix:} 
    $
        \sigma_1 . \sigma_2 \in LUT  \Rightarrow  \sigma_1 \in LUT
    $

    $-$ {\em quiescence:} for any $d<delay(ev)$,
    $
        \sigma_1 .ev \in LUT  \Rightarrow  \sigma_1  . (d, \delta!) \in LUT
    $

    $-$ {\em input complete:}  for any $d<delay(ev)$, $c \in CC^{in}$, $v \in M$,
    
    \centerline{ $
        \sigma_1 . ev \in LUT  \Rightarrow  \sigma_1 . (d, c?v) \in LUT
    $}

The first hypothesis simply states that every prefix of a system trace is also a system trace. The hypothesis on quiescence states that if an event $ev$ whose action is in $act(C)$ occurs in $LUT$, then $LUT$ can remain quiescent for any duration strictly less than the delay of the event. The hypothesis on input completeness enables $LUT$ to receive any input on a controllable channel, i.e., an input received from the test case, during the delay of any $ev$ in the $LUT$.



The semantics of a TIOSTS $\mathbb{G}$, denoted by $Sem(\mathbb{G})$, will include traces with the admissible temporary observation of quiescence: if an event $ev =(d,act)$ is specified in $\mathbb{G}$ then quiescence can be observed for any duration $d' < d$. $Sem(\mathbb{G})$ is then defined as the smallest set containing ${Traces}(\mathbb{G})$ and such that for any $\sigma \in Evt(C)^*$, $ev\in Evt(C)$, for any $d<delay(ev)$:\\
\centerline{$
    \sigma . ev \in Traces(\mathbb{G})  \Rightarrow  \sigma . (d,\delta!) \in Sem(\mathbb{G})
$}



As other previous works \cite{LuthmannGL19,TretmansJ22} have already done to suit their needs, we are now slightly adapting the conformance relation of~\cite{KrichenT04}:

\begin{definition}[tioco]
Let $C$ be a set of channels. Let $\mathbb{G}$ and $LUT$ be resp. a TIOSTS defined on $C$ and a subset of $Evt(C_\delta)^*$ satisfying $\cal{H}$. \\
LUT tioco $\mathbb{G}$ iff for all $\sigma \in Sem(\mathbb{G})$, for any $ev\in Evt(C^{out} \cup \{\delta\})$, we have:
\vspace{-.275cm}
\[
    \sigma . ev \in LUT  \Rightarrow  \sigma . ev \in Sem(\mathbb{G})
\]
\end{definition}
The relation $tioco$ states that $LUT$ is in conformance with $\mathbb{G}$, if and only if after a specified sequence $\sigma$ observed on $LUT$, any event produced by $LUT$ as an output or a delay of quiescence, leads to a sequence $\sigma.ev$ of $sem(\mathbb{G})$.


Test case generation is often based on the selection of a test purpose which permits to choose a particular behavior in $\mathbb{G}$ to be tested~\cite{GastonGRT06,HesselLMNPS08,BannourEGG12,TretmansTap19}. As symbolic execution plays a key role both for the semantics of TIOSTS and for testing issues in general, whether it is for the test case generation or the verdict computation, our test purposes will be paths $tp\in Paths(SE(\mathbb{G})_\delta)$ with satisfiable path conditions, i.e. verifying 
$Sat(tgt(tp))$. As outputs are involved in the $tioco$ relation, we require $tp$ to end with an output event, i.e., $chan(tgt(tp))\in C^{out}$. 
 
Contrary to~\cite{GastonGRT06}, our test purposes are simple paths and not (finite) subtrees, simplifying the construction of test cases. We will not need to consider the case where the observed behavior on $LUT$ corresponds to several symbolic paths simultaneously. By taking it a step further, to avoid such tricky situations completely, we restrict ourselves to symbolic paths that do not induce non-determinism. In line with~\cite{AndradeMJM11,Jeron09}, we forbid there are two outgoing transitions of an execution context concerning the same channel which can be covered by the same trace.
Unlike~\cite{AndradeMJM11,Jeron09} which impose determinism conditions at the state level, we deal with uninitialized variables at the path level:
 
\begin{definition}[Test purpose]\label{def:Test purpose} Let $tp \in Paths(SE(\mathbb{G})^\delta)$ be a symbolic path verifying $Sat(tgt(tp))$ and $chan(tgt(tp))\in C^{out}$. Let $\mathbb{EC}(tp)$ be its set of execution contexts.  

\noindent 
$tp$ is a \emph{test purpose} for $\mathbb{G}$ if $tp$ satisfies the so-called \emph{trace-determinism property}:

\noindent for $ec$ in $\mathbb{EC}(tp)$ and $ec'$ in $\mathbb{EC}(\mathbb{G})$ s.t. $Sat(ec')$, $pec(ec) = pec(ec')$, $tr(ec) \neq tr(ec')$, and $chan(ec) = chan(ec')$, the following formula is unsatisfiable:

\vspace{-.275cm}
\[\big(\exists {F^{ini}}.\pi(ec)\big) \bigwedge \big(\exists {F^{ini}}.\pi(ec') \big)\]
\end{definition}
The trace-determinism property simply expresses that from any intermediate execution context of $tp$, it is impossible to deviate in $\mathbb{G}$ with a common trace, independently of the initial conditions.
\begin{example}\label{ex:tiosts-determ} The test purpose $tp=ec_0.ec_1 \ldots ec_4$ given in Fig.~\ref{fig:tiosts-symbex} satisfies trace-determinism. To support our comments, let us consider a simpler TIOSTS with three transitions ($tr_1$, $tr_2$ and $tr_3$), with two of them, $tr_2$ and $tr_3$, creating a non-deterministic situation:

\noindent 
$tr_1 : q_0 \xrightarrow[]{ \actIn{Transc}  amt } q_1$, 
$tr_2 : q_1 \xrightarrow[]{ \actOut{Debit} amt } q_2$ and 
$tr_3 : q_1 \xrightarrow[]{ [\; fee > 0\; ] \,,\, \actOut{Debit} amt+fee } q_3$ 

\noindent The TIOSTS symbolic execution for some initial execution context 
$ec_0$ ($ F^{ini} = \{ \f{amt_0}, \f{fee_0} \}$) can reach execution contexts 
$ec_1$ ($tr(ec_1)=tr_1$, $pec(ec_1)=ec_0$),  $ec_2$ ($tr(ec_2)=tr_2$, $pec(ec_2)= ec_1$), and  
$ec_3$ ($tr(ec_3)=tr_3$, $pec(ec_3) = ec_1$), building 2 symbolic paths $ec_0.ec_1.ec_2$ and $ec_0.ec_1.ec_3$. Respective path conditions are $ \pi(ec_2) = (\f{\freshOut{1}{D}{}} = \f{amt_1}) $ and $ \pi(ec_3) = (\f{fee_0} > 0) \land (\f{\freshOut{1}{D}{}} = \f{amt_1} + \f{fee_0})$ where $\f{amt_1}$ binds the value received on the channel $\text{Transc}$ ($f^{in}_{\text{Transc}}(ec_0)=\f{amt_1}$) and $\f{ \freshOut{1}{D}{} }$ binds the 
value emitted on the channel $\text{Debit}$  ($f^{out}_{\text{Debit}}(ec_1) = \f{ \freshOut{1}{D}{} }$).

Given $tp=ec_0.ec_1.ec_2$, execution contexts $ec_2$ and $ec_3$ share the same output channel $\text{Debit}$ and the same predecessor context $ec_1$. $tp$ satisfies the trace-determinism property. Indeed, the formula: \\ 
{\small
\centerline{
    $\big( \exists \{\f{fee_0},\f{amt_0}\} . (\f{ \freshOut{1}{D}{} } \!=\! \f{amt_1}) \big)
    \bigwedge  
    \big( 
        \exists \{\f{fee_0},\f{amt_0}\} . (\f{fee_0} \!>\! 0) 
        \wedge 
        (\f{ \freshOut{1}{D}{} } \!=\! \f{amt_1} \!+\! \f{fee_0}) 
    \big)$
}
}
\noindent 
is not satisfiable because $\f{amt_1} < \f{amt_1} + \f{fee_0}$ holds as we have $\f{fee_0} > 0$. A trace cannot belong to distinct paths: if the debit value is the same as what is requested for withdrawal then trace covers $ec_2$, else 
$ec_3$ is covered. 
\end{example}

\vspace{-.25cm}
\section{Path-guided test cases}
\label{sec:testcase}

Roughly speaking, a test case $\mathbb{TC}$ will be a mirror TIOSTS of a TIOSTS $\mathbb{G}$, restricted by $tp$, a test purpose of $\mathbb{G}$, intended to interact with a $LUT$ that we wish to check its conformance to $\mathbb{G}$ up to $tp$.
$\mathbb{TC}$ will be a tree-like TIOSTS with $tp$ of $\mathbb{G}$ as a backbone, incorporating the following specific characteristics:
\begin{itemize}
    \item execution contexts in $\mathbb{EC}(tp)$ constitute the main branch,
    \item sink states or leaves are assimilated with a test verdict. Notably, the last execution context $tgt(tp)$ of $tp$ will be assimilated with the $\text{PASS}$ verdict,
    \item from each $ec$ in $\mathbb{EC}(tp)$ other than $tgt(tp)$, the outgoing arcs outside $tp$ decline all ways to deviate from $tp$ and directly lead to a verdict state, either an inconclusive verdict or a failure verdict.
\end{itemize}

In Def.\ref{def:test-case}, we define $\mathbb{TC}$ by enumerating the different cases of transitions to be built according to channel type and $tp$ membership. We now give a few indications for enhancing the definition readability. Channel roles are reversed: channels in $CC^{in}$ as well as 
channel $\delta$ (resp. $C^{out} \cup UC^{in}$) become output (resp. input) channels. 
Variables of $\mathbb{TC}$ will be 
 symbolic variables involved in $tp$ and will be used to store successive concrete events observed on $LUT$. Any observation  on $LUT$ will be encoded as an input transition in $\mathbb{TC}$, whether it is an emission from a channel of $C^{out}$, a reception on a channel of $UC^{in}$ or a time-out observation. On the latter, it is conventional to consider that a system that does not react before a certain delay, chosen to be long enough, is in a state of quiescence. The only notable exception is when input transitions of $tp$ give rise to output transitions for $\mathbb{TC}$, modeling a situation in which $\mathbb{TC}$ stimulates $LUT$ by sending it data. The choice of the data to send is conditioned by two constraints: (i) taking into account the information collected so far and stored in the variables of $\mathbb{TC}$ in the first steps, and (ii) the guarantee of being able to reach the last EC of $tp$, i.e. the verdict $\text{PASS}$. This will be done by ensuring the satisfiability of the path condition of $tp$, leaving aside the variables already binded. 

We will refer to variables of a symbolic path as follows: for\footnote{For $i$ and $j$ in $\mathbb{N}$ verifying $i < j$, $[i,j)$ contains the integers from $i$ to $j-1$ included.} $\alpha \in \{in,out\}$, $f^\alpha(p)=\bigcup_{i \in [1,n)} f^\alpha(ec_i)$ will denote all introduced fresh variables in $F^\alpha$ used to compute a symbolic path $p =ec_0.ec_1\ldots ec_n$. Similarly, $f^{dur}(p) = \{ f^{dur}(ec_i) | i \in [1,n) \}$. Moreover, for a target execution context $ec_n = tgt(p)$, we denote $\overline{f}^\alpha(ec_n)$ and $\overline{f}(ec_n)$ resp. for $f^\alpha(p)$ and  $f(p)$. 

\begin{definition}[Path-guided test case]
\label{def:test-case}
Let $tp$ be a test purpose for a TIOSTS $\mathbb{G}$. 
Let us consider the signature $\widehat{\Sigma} = (\widehat{A}, \widehat{K}, \widehat{C})$ where: 
\begin{itemize}
    \item $\widehat{A} = f^{in}(tp) \cup f^{out}(tp)$, 
    \item $\widehat{K}=f^{dur}(tp)$, 
    \item $\widehat{C}=C_\delta$ such that $\widehat{C}^{in} = C^{out} \cup UC^{in}$ and $\widehat{C}^{out} = CC^{in} \cup \{\delta\}$
\end{itemize}    
A \emph{test case guided by} $tp$ is a TIOSTS $\mathbb{TC}=(\widehat{Q}, \widehat{q_0}, \widehat{Tr})$ over $\widehat{\Sigma}$ where:  
\begin{itemize}
    \item $\widehat{Q} = \big(\mathbb{EC}(tp) \setminus \{tgt(tp)\} \big) \cup \mathbb{V}$ where: \\
    $\mathbb{V} = \{$
    $\emph{PASS},$
    $\emph{FAIL}^{out},$
    $\emph{FAIL}^{dur},$
    $\emph{INC}^{out},$
    $\emph{INC}^{dur},$
    $\emph{INC}^\uncIn_\spec,$
    $\emph{INC}^\uncIn_\unspec$
$\}$,
    \item $\widehat{q_0}=ec_0$,
    \vspace{-.2cm}
    \item $\widehat{Tr}$ is defined by a set ${\cal R}$ of  10 rules of the form \resizebox{!}{20pt}{
$
\infer[\begin{array}{l}(Ri) \\ \emph{LABEL} \end{array}]
     {tr\!\in\!\widehat{Tr}}
     {H}$
    } 
    for $i \in [1,10]$. Such a rule reads as follows: the transition $tr$ can be added due to rule $Ri$ to $\widehat{Tr}$ provided that hypothesis $H$ holds and $Sat(\phi(tr))$.
\end{itemize}   

In writing the rules of ${\cal R}$, we will use the following formulas

\resizebox{!}{59pt}{
$\begin{array}{lcl}
&\\
    \phi_{stim} &:&
    \exists F^{ini} \cup \overline{f}(tgt(tp)) 
    \setminus \overline{f}(ec'). \pi(tp) 
\\
&
\\
\vspace{.2cm}
\phi^\obs_\spec &:&
    f^{dur} (ec) < \emph{TM}
    \wedge (\exists F^{ini}.\pi(ec'))
\\
\phi^\obs_\unspec &:&
        f^{dur} (ec) < \emph{TM}  \land  \bigwedge_{\substack{pec(ec') = ec \\ chan(ec')=c }}(\forall F^{ini}.\neg\pi(ec'))
\\
&
\\
\phi^{\delta}_\spec &:& 
f^{dur}(ec) \geq \emph{TM}
    \land \bigvee_{\substack{pec(ec')=ec 
        \\ chan(ec') \in C^{out} \cup UC^{in} \cup \{\delta\}}} (\exists F^{ini}.\pi(ec'))
\\
\phi^{\delta}_\unspec &:& 
f^{dur}(ec) \geq \emph{TM}
    \land \bigwedge_{\substack{pec(ec')=ec 
        \\ chan(ec') \in C^{out} \cup UC^{in} \cup \{\delta\}}} (\forall F^{ini}.\neg \pi(ec'))
\\
&
\end{array}
$ }

where the constant $\emph{TM}$ (Time-out) 
sets the maximum waiting-time for observing outputs or uncontrollable inputs. 

\vspace{.2cm}
{
\noindent\resizebox{!}{25pt}{
$
\!\!\!\infer[\!\!\begin{array}{l}(R1) \\ \emph{SKIP} \end{array}]
     {\big(
      ec
, c!x
, \phi_{stim}
, \{ f^{dur} (ec) \}
, id_{\widehat{A}}
, ec'
\big)\!\in\!\widehat{Tr}}
     {
     ec\xrightarrow[]{(z, c?x)} ec' \in \mathbb{EC}(tp) \qquad c \in CC^{in}
     }
$
}

\vspace{.4cm}

\noindent\resizebox{!}{35pt}{
$
\!\!\!\infer[\!\!\begin{array}{l}(R2) \\ \emph{SKIP} \end{array}]
{
\big( ec
, c?y
, \phi^\obs_\spec
, \{ f^{dur} (ec') \}
, id_{\widehat{A}}
, ec' \big)\!\in\!\widehat{Tr}
}
{
\begin{array}{l}
ec\xrightarrow[]{(z, c!y)} ec' \in \mathbb{EC}(tp)
\\
ec' \neq tgt(tp)\qquad c \in C^{out}
\end{array}
}
\quad
\infer[\!\!\begin{array}{l}(R3) \\ \emph{PASS} \end{array}]
{
\big(
     ec
, c?y
, \phi^\obs_\spec
, \emptyset
, id_{\widehat{A}}
, \emph{PASS} \big)\!\in\!\widehat{Tr}
}
{
\begin{array}{l}
     ec\xrightarrow[]{(z, c!y)} ec' \in \mathbb{EC}(tp)
     \\
    ec' = tgt(tp)\qquad c\in C^{out}
\end{array}
}
$
}

\noindent\resizebox{!}{35pt}{
$
\!\!\!\infer[\!\!\begin{array}{l}(R4) \\ \emph{INC}^{out} \end{array}]
{
\big(
      ec
, c?y
, \phi^\obs_\spec
, \emptyset
, id_{\widehat{A}}
, \emph{INC}^{out}\big)\!\in\!\widehat{Tr}
}
{
\begin{array}{l}
     ec\xrightarrow[]{(z, c!y)} ec' \;\; ec\in \mathbb{EC}(tp)\setminus \{tgt(tp)\} 
\\
    ec' \not\in\mathbb{EC}(tp)\qquad c\in C^{out}
\end{array}
}
\quad
\infer[\!\!\begin{array}{l}(R5) \\ \emph{FAIL}^{out} \end{array}]
{
\big(
      ec
, c?f^{out}_c(ec)
, \phi^\obs_\unspec
, \emptyset
, id_{\widehat{A}}
, \emph{FAIL}^{out}\big)\!\in\!\widehat{Tr}
}
{
\begin{array}{l}
     \textcolor{white}{ec\xrightarrow[]{(z, c!y)} ec'
\;\; ec \in \mathbb{EC}(tp) \setminus \{tgt(tp)\}}
\\
    ec \in \mathbb{EC}(tp) \qquad c \in C^{out}
\end{array}
}
$
}

\vspace{.425cm}

\noindent\resizebox{!}{35pt}{
$
\!\!\!\infer[\!\!\begin{array}{l}(R6) \\ \emph{SKIP} \end{array}]
{
\big(
      ec
, c?x
, \phi^\obs_\spec
, \{ f^{dur} (ec') \}
, id_{\widehat{A}}
, ec' \big)\!\in\!\widehat{Tr}
}
{
\begin{array}{l}
     ec\xrightarrow[]{(z, c?x)} ec' \in  \mathbb{EC}(tp)
     \\
     c \in UC^{in}
     \end{array}
}
\quad 
\infer[\!\!\begin{array}{l}(R7) \\ \emph{INC}^\uncIn_\spec  \end{array}]
{
\big(
      ec
, c?x
, \phi^\obs_\spec
, \emptyset
, id_{\widehat{A}} 
, \emph{INC}^\uncIn_\spec\big)\!\in\!\widehat{Tr}
}
{   \begin{array}{l}
     ec\xrightarrow[]{(z, c?x)} ec' \;\; ec \in \mathbb{EC}(tp) \setminus \{tgt(tp)\}
\\
    ec' \not\in \mathbb{EC}(tp) \qquad c \in UC^{in}
    \end{array}
}
$
}

\vspace{.425cm}
\noindent\resizebox{!}{20.5pt}{
$
\!\!\!\infer[\!\!\begin{array}{l}(R8) \\ \emph{INC}^\uncIn_\unspec \end{array}]
     {\big(
      ec
, c?f^{in}_c(ec)
, \phi^\obs_\unspec
, \emptyset
, id_{\widehat{A}} 
, \emph{INC}^\uncIn_\unspec\big)\!\in\!\widehat{Tr}}
     {
     ec \in \mathbb{EC}(tp) \qquad c \in UC^{in}
     }
$
}

\vspace{.425cm}

\noindent\resizebox{!}{20pt}{
$
\!\!\!\infer[\!\!\begin{array}{l}(R9) \\ \emph{INC}^{dur}\end{array}]
     {\big(
      ec
, \delta!
, \phi^{\delta}_\spec
, \emptyset
, id_{\widehat{A}}
, \emph{INC}^{dur}\big)\!\in\!\widehat{Tr}}
     {
     ec \in \mathbb{EC}(tp)
     }
\quad
\infer[\!\!\begin{array}{l}(R10) \\ \emph{FAIL}^{dur}\end{array}]
     {\big(
      ec , \delta!
, \phi^\delta_\unspec
, \emptyset
, id_{\widehat{A}} 
, \emph{FAIL}^{dur}\big)\!\in\!\widehat{Tr}}
     {
     ec \in \mathbb{EC}(tp)
     }
$
}

}

\end{definition}

A verdict $\text{PASS}$ is reached when $tp$ is covered, verdicts $\text{INC}^m_n$ are reached when traces deviate from $tp$ while remaining in $\mathbb{G}$, and verdicts $\text{FAIL}^m$ denote traces outside $\mathbb{G}$. The annotations $n$ and $m$ provide additional information on the cause of the verdict. Rules $R1$, $R2$ and $R6$, grouped together under the label $\text{SKIP}$, allow advancing along $tp$, resp. by stimulating $LUT$ with the sending of data, observing an emission on $C^{out}$ and observing a reception on $UC^{in}$. Rule $R3$ 
indicates that the last EC of $tp$, thus the $\text{PASS}$ verdict, has been reached. Rules $R4$, $R7$, $R8$ and $R9$, each with a label $\text{INC}^m_n$ indicate that the 
observed event causes $LUT$ to leave $tp$, without leaving $\mathbb{G}$, resp. by observing an output, an input specified in $\mathbb{G}$, an input not specified in $\mathbb{G}$ and a time-out observation. Lastly, rules $R5$ and $R10$, resp. labeled by  $\text{FAIL}^{out}$ and $\text{FAIL}^{dur}$, raise a $\text{FAIL}$ verdict, for resp. an unauthorized output and an exceeded time-out.

\begin{figure}[t] 
\vspace{-.275cm}
    \centering
    
    \resizebox{1\textwidth}{!}{
        \hspace{-1cm}\input{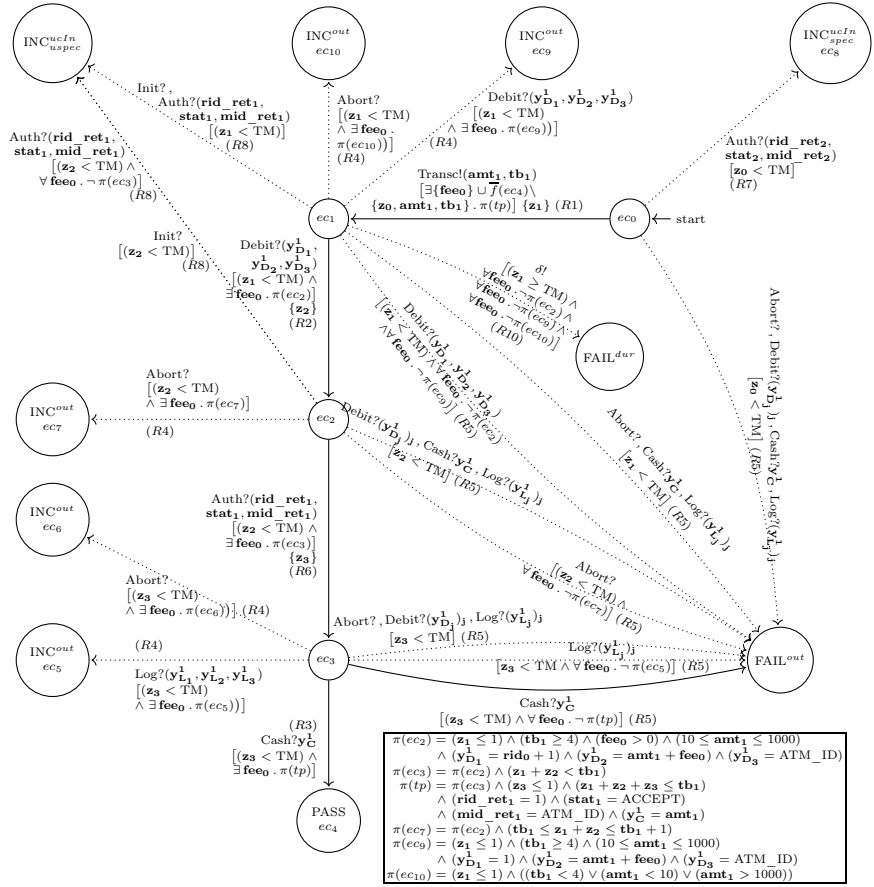}
    }
	\caption{Test case for $\text{ATM}$}
	\label{fig:test_case_ATM}
 \vspace{-.6cm}
\end{figure}

\begin{example}
\label{ex:test-case} In Fig.~\ref{fig:test_case_ATM}, the test case for $tp=ec_0ec_1\ldots ec_4$ (see Ex.~\ref{ex:tiosts-determ}) is depicted. Certain verdict states are repeated for readability, and defining rules annotate the transitions. The test case utilizes the $\text{Transc}$ channel as a controllable input channel for stimulation while observing all other channels. For space considerations, we comment only on some rules.

 \noindent Rule $R1$ defines a stimulation action $\actOut{Transc}(\f{amt_1},\f{\vartime_1})$ (transition from $ec_0$ to $ec_1$) constrained by $\pi(tp)$ to select an appropriate value for $\f{amt_1}$ and $\f{\vartime_1}$ together with a time of stimulation $\f{z_0}$ that allows to follow the test purpose. Within $\pi(tp)$, $\f{amt_1}$ is limited to some range ($10\leq \f{amt_1} \leq 1000$), $\f{\vartime_1}$ is greater than or equal $4$, while $\f{z_0}$ is unconstrained (a duration measured with clock $\f{z_0}$). Non-revealed variables, i.e., other than $\f{z_0}$, $\f{amt_1}$ and $\f{\vartime_1}$ appearing in $\pi(tp)$ are bound by the existential quantifier due to their unknown values at this execution point. The clock $\f{z_1}$ is reset to enable reasoning on subsequent actions' duration (measured on $\f{z_1}$). Rule $R2$ defines an observation action $\actIn{Debit}(\f{\freshOut{1}{D}{1}},\f{\freshOut{1}{D}{2}},\f{\freshOut{1}{D}{3}})$  constrained by 
$(\f{z_1}\!<\!\maxWaitingBound) 
\wedge \exists \f{fee_0} . \piAlias{2}$ (transition from $ec_1$ to $ec_2$) while rule $R5$ defines the same observation constrained by 
 $(\f{z_1}\!<\!\maxWaitingBound)
    \land \forall \f{fee_0} .
    \neg \, \piAlias{2}
    \land \forall \{ \f{fee_0} \} \, . \,
    \neg \, \piAlias{9}
    $ (transition from $ec_1$ to $\text{FAIL}^{out}$). Both situations are possible: Trace $(0,\actOut{Transc}(50,4)).
    (0,\actIn{Debit}(1,51,\ATMID))$ reaches $ec_2$ whereas $\text{FAIL}_{out}$ is reached by traces $(0,\actOut{Transc}(50,4)).
    (0,\actIn{Debit}(1,0,\ATMID))$ and $(0,\actOut{Transc}(50,4)).
    (2,\actIn{Debit}(1,51,\ATMID))$ due resp. to data and time non-compliance. Rule $R6$ defines an unspecified quiescence $\delta!$ constrained by $(\f{z_1} \geq \maxWaitingBound)
    \land 
    \forall \f{fee_0} \, . \neg \piAlias{2}
    \land
    \forall \f{fee_0} \, . \neg \piAlias{9}
    \land
    \forall \f{fee_0} \, .\neg \piAlias{10}
$ (transition from $ec_1$ to $\text{FAIL}^{dur}$). The trace $(0,\actOut{Transc}(50,4)).(5,\delta!)$ reaches $\text{FAIL}^{dur}$ (time-out $\maxWaitingBound$ is set to $5$): the time-out is exceeded without the mandatory output on the channel $\text{Debit}$ being observed. Rule $R9$ defines a specified quiescence $\delta!$ (transition from $ec_1$ to $\text{INC}^{dur}$ not drawn for space). This case arises when there is still sufficient time to reach $ec_2$, $ec_9$ or $ec_{10}$ (no quiescence applies from $ec_1$, see Ex.~\ref{ex:quiescence}).
\end{example}
\vspace{-.275cm}

A test case  $\mathbb{TC}$ interacts with a $LUT$, designed to comply with a TIOSTS $\mathbb{G}$, to issue a verdict about a test purpose $tp$. $\mathbb{TC}$ is therefore defined as a mirror image of $\mathbb{G}$: emissions (receptions) in $\mathbb{TC}$ correspond to receptions (emissions) of $\mathbb{G}$ (cf. rules $R1$ and $R2$), except uncontrollable channels whose actions are not reversed (cf rule $R6$). Given a concrete action $act$ (with $v$ for value received or sent), we denote $\overline{act}$ its mirror action, defined as follows: $\overline{c!v} = c?v$ for $c \in C^{out}$, $\overline{c?v} = c!v$ for $c \in CC^{in}$ and  $\overline{c?v} = c?v$ for $c \in UC^{in}$.
We introduce an execution relation that abstracts a synchronized execution of a trace with $\mathbb{TC}$: 
\begin{definition}[Relation execution $\leadsto_{\mathbb{TC}}$]
\label{def:relation-exec}
Let $\mathbb{G}$ be a TIOSTS and $tp$ a test purpose for $\mathbb{G}$ with $\mathbb{TC} = (\widehat{Q},\widehat{q_0},\widehat{Tr})$ the test case guided by $tp$.

The \emph{execution relation}  $\leadsto_{\mathbb{TC}}\subseteq \big( Evt(C_\delta)^* \times  \widehat{Q} \times M^F \big)^2$  is defined by:

for $ev.\sigma \in Evt(C_\delta)^*$, $q,q' \in \widehat{Q}$ and for $\nu, \nu' \in M^F$, $(ev.\sigma,q,\nu) ~\leadsto_{\mathbb{TC}}~ (\sigma, q',\nu')$ holds iff there exists $tr\in \widehat{Tr}$ s.t. $src(tr)\!=\! q$, $tgt(tr)\!=\!q'$, $\nu'(act(tr))=\overline{act(ev)}$ and $M \models_{\nu'} \phi(tr)$ with $\nu'$ defined as:
 \begin{itemize}
     \item if $ev\!=\!(d,c?v)$ and $chan(tr)\!=\!c$ then $\nu'\!=\!\nu[f^{dur}(q)\!\mapsto\!d][ f^{in}_c(q)\!\mapsto\!v]$ ;
     \item if $ev\!=\!(d,c!v)$ and $chan(tr)\!=\!c$ then $\nu'\!=\!\nu[f^{dur}(q)\!\mapsto\!d][f^{out}_c(q) \!\mapsto\! v]$ ;
     \item else, i.e., $ev\!=\!(d,\delta!)$ and $chan(tr)\!=\!\delta$, $\nu'=\nu[f^{dur}(q)\!\mapsto\!d]$.
 \end{itemize}
 \end{definition}


Intuitively, a step $(ev.\sigma, q,\nu) ~\leadsto_{\mathbb{TC}}~ (\sigma, q',\nu')$ consists in:
\begin{itemize}
    \item reading the first element $ev$ of a trace from a test case state $q$ and an interpretation $\nu$  synthesizing the known information about the variables in~$F$;
    \item finding a transition $tr$ in $\widehat{Tr}$ whose action matches the mirror action of $ev$; 
    \item building a new triple with $\sigma$ the trace remaining to be read, $q'$ a successor state of $q$ in $\widehat{Q}$, and $\nu'$ the updated interpretation of the variables~$F$. 
\end{itemize}
The execution relation simulates a parallel composition between timed input output systems, synchronizing inputs and outputs. Our formulation deviates from the one in \cite{KrichenT06} for two essential reasons: the symbolic nature of the test case requires the intermediate interpretations of variables to be memorized, and uncontrollable channels require to adapt the synchronization~\cite{EscobedoGGC10}.
Let $LUT$ be a subset of $ Evt(C_\delta)^*$ satisfying ${\cal H}$ and $\overset{*}{\leadsto}_{\mathbb{TC}}$ be the reflexive and transitive closure of  $\leadsto_{\mathbb{TC}}$. Given a $LUT$ trace $\sigma_0$, we apply the execution relation iteratively from an initial triplet consisting of $\sigma_0$ a trace, $ec_0$ an initial EC and $\nu_0$ an arbitrary interpretation, to obtain the corresponding test verdict for $tp$, so that the verdict set obtained from the execution of $\mathbb{TC}$ on $LUT$ is defined by: 


\noindent $vdt(LUT,\mathbb{TC})= \{\text{V} ~|~ \exists \sigma_0 \in LUT, \nu_0 \in M^F, (\sigma_0,{q_0}_{\tc}, \nu_0) ~\overset{*}{\leadsto}_{\mathbb{TC}} (\sigma,\text{V}, \nu )\}$.

\noindent Th.~\ref{th:corr} states the soundness of the test case execution for detecting errors through the $\text{FAIL}^{out}$ and $\text{FAIL}^{dur}$ verdicts.  A proof can be found in in Appendix~\ref{anx:proof}. Comparable results can be formulated for the other verdicts. Still, those relating to FAIL verdicts are the only ones to guarantee that any discarded system under test does not satisfy the tioco conformance relation.

\begin{theorem}
\label{th:corr} 
Let $C$ be a set of channels. Let $\mathbb{G}$ and $LUT$ be resp. a TIOSTS defined on $C$ and a subset of $Evt(C_\delta)^*$ satisfying $\cal{H}$. \\
 If $LUT~tioco~\mathbb{G}$ then for any test purpose $tp$ for $\mathbb{G}$ with $\mathbb{TC}$ as test case guided by $tp$, we have $\emph{FAIL}^{out}\not\in vdt(LUT, \mathbb{TC}) $ and $\emph{FAIL}^{dur}\not\in vdt(LUT, \mathbb{TC})$.
\end{theorem}


The test case generation is implemented as module in the Diversity symbolic execution platform~\cite{diversity}. Resulting test cases are expressed in Diversity's entry language, allowing their exploration through symbolic execution with the SMT-solver Z3~\cite{Z3}. For easier execution, we export the test cases from Diversity in JSON format, with transition guards expressed in the SMT-LIB input format for SMT-solvers. Our experiments involved applying this test case generation to the ATM example on an Intel Core i7 processor. Varying the size of the test purposes up to $100$ transitions, we observed successful trace-determinism verification for all test purposes. We noted a noticeable increase in generation duration as the test purpose size grew, while still remaining feasible. Generating the TIOSTS test case ($513$ transitions) for the test purpose of $100$ transitions took more than $40$s, in contrast to only $500$ms for the test purpose of size $4$ in Ex.~\ref{ex:test-case} ($31$ transitions) resp. $8$s for the test purpose of size $50$ ($138$ transitions).

\vspace{-.25cm}
\section{Related work} 
\label{sec:RelatedWork}

Existing works for (t)ioco conformance test cases from symbolic models employs two main generation methods: online and offline. Online generation~\cite{FrantzenTW04,GastonGRT06,TretmansTap19} involves dynamically generating test cases 
while exploring the model during execution on the system under test. In contrast, offline generation~\cite{HenryJM22,AndradeMJM11,BannourEGG12} focuses on deriving test cases from the model before executing them on the system under test. Some works \cite{FrantzenTW04,GastonGRT06,TretmansTap19} propose online test case generation using symbolic execution. Yet, these works did not consider time constraints, and in particular, work~\cite{TretmansTap19} did not consider quiescence. In~\cite{TretmansTap19}, a test purpose is a finite path, while in~\cite{GastonGRT06}, it is a finite symbolic subtree. Both works require maintaining a set of reached symbolic states during test case execution to avoid inconsistent verdicts in case of non-determinism, at the expense of computational resources for tracking the symbolic states and solving their path conditions.
Work~\cite{BannourEGG12} proposes offline test case generation using a path to compute a timed stimulation sequence for the system under test. The recorded timed output sequence is then analyzed for conformance. This approach lacks control over the value and timing of the next stimulation relative to the observed system behavior, potentially resulting in greater deviations from the test purpose. In~\cite{HenryJM22}, objective-centered testers for timed automata are built using game theory. 
Works~\cite{Jeron09,AndradeMJM11} propose offline test case generation as tree-like symbolic transition systems and thus restricted to determinism as we do. The test case generation in~\cite{Jeron09} relies on abstract interpretation to reinforce test case guards on data to keep chances of staying in the test purpose and does not consider time. In~\cite{AndradeMJM11}, symbolic execution techniques are used for data handling, while zone-based abstraction techniques are employed for time. This separation results in less expressive and flexible models, as it cannot extend to incorporate data parameters in the time constraints.


\vspace{-.25cm}
\section{Conclusion}
\label{sec:Conclusion}
This paper presents an offline approach to conformance test case generation from models of timed symbolic transition systems using symbolic execution to handle data and time. Our test purpose is a symbolic path in the model which fulfills a determinism condition to enable the generation of sound tree-like test cases. By distinguishing between controllable inputs (from the test case) and uncontrollable inputs (from other systems), our approach enhances the usability of test cases when the system interacts with other systems (remote in general). It should be noted that our test cases include configurations where the resolution time exceeds the stimulation time or overlaps the arrival of an observation.


\bibliographystyle{splncs04}
\bibliography{biblio}

\newpage

\appendix
\section{Proof of Th.~\ref{th:corr}}
\label{anx:proof}

\noindent 
We carry out the proof by contraposition, i.e. we prove that if there exists a test objective $tp$ 
for $\mathbb{G}$ and a test case $\mathbb{TC}$ guided by $tp$ s.t. $\text{FAIL}^{out}$ or $\text{FAIL}^{dur}$ are in $vdt(LUT, \mathbb{TC})$ then $\neg(LUT~tioco~\mathbb{G})$. According to the definition of tioco, this amounts to proving that there exists $\sigma \in Sem(\mathbb{G}) \cap LUT$, and an $ev \in Evt(C^{out} \cup \{\delta\})$ s.t. $\sigma. ev \in LUT$ and $\sigma.ev\not\in Sem(\mathbb{G})$. 
Prop.\ref{prop:corr} states this claim.

\begin{proposition}
\label{prop:corr}
Let $\sigma . ev . \sigma'$ be a trace of $LUT$ with $\sigma$ and $\sigma'$ traces in $Evt(C_\delta)^*$ and $ev$ a concrete event in $Evt(C_\delta)$.
If there exists a test purpose $tp$ for $\mathbb{G}$ and a test case $\mathbb{TC}$ guided by $tp$ s.t. $(\sigma.ev.\sigma',ec_0, \nu_0) ~\overset{*}{\leadsto}_{\mathbb{TC}} (\sigma',\emph{V}, \nu' )$ with $\emph{V} \in \{\emph{FAIL}^{out},\emph{FAIL}^{dur}\}$ then:
\begin{enumerate}

\item[{(1)}] $ev\in Evt(C^{out} \cup \{\delta\})$

\item[{(2)}] $\sigma \in Sem(\mathbb{G})$ 

\item[{(3)}] $\sigma.ev\not\in Sem(\mathbb{G})$
\end{enumerate}
\end{proposition}

\noindent 
Let us suppose $(\sigma.ev.\sigma',ec_0, \nu_0) ~\overset{*}{\leadsto}_{\mathbb{TC}} (\sigma',\emph{V}, \nu' )$ with $\text{V} \in \{\text{FAIL}^{out},\text{FAIL}^{dur}\}$. There exists an execution sequence  $(\sigma.ev.\sigma',ec_0, \nu_0) \overset{*}{\leadsto}_{\mathbb{TC}} (ev.\sigma',ec,\nu){\leadsto}_{\mathbb{TC}} (\sigma',\text{V}, \nu' )$ s.t. $ev$ is executed with a Test-Case transition (TC-transition) $tr$ built with either rule $R5$ $\text{FAIL}^{out}$ or rule $R10$ $\text{FAIL}^{dur}$.

\medskip

{\bf{\em Proof of (1)}}  As $chan(tr)=chan(ev)$, by  $R5$ $\text{FAIL}^{out}$ (resp. $R10$ $\text{FAIL}^{dur}$), $chan(tr) \in C^{out}$ (resp. $chan(tr)=\delta$). Hence $ev\in Evt(C^{out} \cup \{\delta\})$.

{\bf{\em Proof of (2)}} Based on the execution steps and corresponding TC-transitions, the events in $\sigma$ can only be of the form $(d,c!v)$ or $(d,c?v)$, excluding the form $(d,\delta!)$. Indeed, no TC transitions on the channel $\delta$ lead to an execution context. The only TC-transitions involving channel $\delta$ are from rules $R9$ $\text{INC}^{dur}$ and $R10$ $\text{FAIL}^{dur}$, and their target state is a verdict.  

Let $ec$ an execution context of $tp$ distinct from $tgt(tp)$. Following its predecessor contexts, it defines a unique path $p_{ec}$ of $\mathbb{G}$ leading to $ec$ defining a non-empty set of traces. Let $\nu$ in $M^F$ an interpretation verifying $M \models_\nu \exists F^{ini}. \pi(tgt(p_{ec}))$ and let $\sigma_\nu$ be the trace of $p_{ec}$ built using $\nu$.

Let us now consider $(ev.\sigma,ec,\nu)  \leadsto_{\mathbb{TC}} (\sigma,ec',\nu')$ with $ec \not\in V$. Let us denote $tr$ the transition of $\widehat{Tr}$ underlying the execution step $(ev.\sigma,ec,\nu)  \leadsto_{\mathbb{TC}} (\sigma,ec',\nu')$. We have $tgt(tr) = ec'$ and $src(tr) = ec$. As $ec \not\in V$, the rule involved in building the transition $tr$ is one of the $R1$, $R2$ and $R6$ rules.



\begin{itemize}
    \item In case $ev=(d,c!v)$, $tr$ comes from the use of rule $R2\;\text{SKIP}$.
    $ev(ec')=(z,c!y)$ with $z=f ^{dur}(ec')$ and $y=f^{out}_c(ec')$; and $\phi(tr) = (z < \maxWaitingBound) \land (\exists F^{ini}.\pi( ec') )$. On the other hand $M \models_{\nu'}  \phi(tr)$ where $\nu'=\nu[z \mapsto d][y\mapsto v]$ (Def.\ref{def:relation-exec} of 
    $\leadsto_{\mathbb{TC}}$) then $M \models_{\nu'} \exists F^{ini}.\pi(ec)$ ($\nu'$ satisfies the right-side of the conjunction of $\phi(tr)$). We can thus extend the trace $\sigma_{\nu}$ of $p_{ec}$ with the concrete event $(d,c!v) = \nu'(ev(ec'))$ to constitute a trace of $p_{ec}.ec'$.  Then $\sigma_{\nu}.(d,c!v) \in Sem(\mathbb{G})$. 
\item 
In case $ev=(d,c?v)$ with $c\in UC^{in}$, $tr$ comes from the use of rule $R6\;\text{SKIP}$. Similar reasoning as for 
 case $ev = (d,c!v)$ ($R2\;\text{SKIP}$) applies. 
\item 
In case $ev=(d,c?v)$ with $c\in CC^{in}$, $tr$ comes from the use of rule $R1\;\text{SKIP}$. We have $tgt(tr)=ec'$; $ev(ec')=(z,c?x)$ with $z=f ^{dur}(ec)$ and $x=f^{in}_c(ec)$ and $\phi(tr)=\exists F^{ini} \cup \overline{f}(tgt(tp))  \setminus \overline{f}(ec'). \pi(tp)$. On the other hand $M \models_{\nu'}  \phi(tr)$ where $\nu'=\nu[z \mapsto d][x\mapsto v]$. As $ec'$ belongs to $\mathbb{EC}(tp)$, path condition $\pi(ec')$ occurs in the conjunction defining formula $\pi(tp)$, then $M \models_{\nu'} \exists F^{ini} \cup \overline{f}(tgt(tp))  \setminus \overline{f}(ec').\pi( ec')$. Also $\nu'$ satisfies $\pi(ec')$ for the values $d$ and $v$  associated with $f^{dur}(ec)$ and $f^{in}_c(ec)$ because these variables are excluded from the variables which are bound by the existential quantifier in $\phi(tr)$. The rest of variables of $\overline{f}(tgt(tp))$ are irrelevant as they dot not occur in $\pi(ec')$, then $M \models_{\nu'} \exists F^{ini}.\pi( ec')$. We can thus extend the trace $\sigma_\nu$ of $p_{ec}$ with the concrete event  $(d,c?v)$ so that $\sigma_\nu. (d,c?v) \in Sem(\mathbb{G})$.
\end{itemize} 
Thus, moving along $tp$ without reaching the $tgt(tp)$ provides traces of $\mathbb{G}$.

\medskip

{\bf{\em Proof of (3)}} Given the execution $(\sigma.ev.\sigma',ec_0, \nu_0) ~\overset{*}{\leadsto}_{\mathbb{TC}} (\sigma',\text{V}, \nu' )$ with $\text{V} \in \{\text{FAIL}^{out},\text{FAIL}^{dur}\}$, we prove that $\sigma.ev\not\in Sem(\mathbb{G})$. 

Let us remark that the path $p$ leading to $ec$ ($tgt(p)=ec$) admits traces $\sigma=\nu(p)$ for interpretations verifying $\nu \models \exists F^{ini}.\pi(ec)$. Moreover, given such a trace $\sigma$, by Prop.\ref{prop:corr} (2), $\sigma \in Sem(\mathbb{G})$ and trace-determinism makes such a corresponding path $p$ unique.
\medskip 

\noindent 
$\bullet$ Case $\text{FAIL}^{out}$ (rule $R5\;\text{FAIL}^{out}$): by Prop.\ref{prop:corr} (1), $ev$ is of the form $(d,c!v)$. 
For the TC-transition $tr$ that allows the execution of $ev$, $chan(tr)=chan(ev)$, $src(tr)=ec$, $tgt(tr)=\text{FAIL}^{out}$ and 

\noindent $\begin{array}{l}
     \phi(tr) =  (f^{dur} (ec)  < \maxWaitingBound )   \land  \bigwedge_{\substack{pec(ec') =ec \\ chan(ec')=c }}(\forall F^{ini}.\neg\pi(ec'))
\end{array}
$ 

\noindent On the other hand $\nu'\models \phi(tr) $ with $\nu'=\nu[f^{dur}(ec)\mapsto d][f^{out}_c(ec) \mapsto v]$ thus $\nu' \models \forall F^{ini}.\neg\pi(ec')$ for any successor $ec'$ of $ec$ such that $chan(ec')=c$. That is, for any such successor $ec'$, $\nu' \not\models \exists F^{ini}.\pi(ec')$. Thus, there is no $ec'$ which extends $p$ into a path $p.ec'$ carrying $\sigma.ev$ through $\nu'$, i.e., such that $\nu' \models \exists F^{ini}.\pi(ec')$ and $\nu'(p.ec')=\sigma.ev$. Then $\sigma.ev\not\in Traces(\mathbb{G})$. 

\noindent $\bullet$ Case $\text{FAIL}^{dur}$ (rule $R10\;\text{FAIL}^{dur}$): 
by Prop.\ref{prop:corr} (1), $ev$ is of the form $(d,\delta!)$. 
For the the TC-transition $tr$ that allows the execution of $ev$, $chan(tr)=chan(ev)$, $src(tr)=ec$, $tgt(tr)=\text{FAIL}^{dur}$ and

\noindent $\begin{array}{l}
      \phi(tr) = f^{dur}(ec) \geq \maxWaitingBound
    \land \bigwedge_{\substack{pec(ec')=ec 
        \\ chan(ec') \in C^{out} \cup UC^{in} \cup \{\delta\}}} \forall F^{ini}.\neg \pi(ec')
\end{array}
$ 

\noindent On the other hand, $\nu'\models \phi(tr) $ with $\nu'=\nu[f^{dur}(ec)\mapsto d]$ thus $\nu' \models \forall F^{ini}.\neg\pi(ec')$ for any successor $ec'$ of $ec$ (i.e. verifying $pec(ec')=ec$) such that $chan(ec')\in C^{out} \cup UC^{in} \cup \{\delta\}$.  If $chan(ec')=\delta$, there is at most one of such successor $ec'$ by the quiescence enrichment. Yet as $\nu' \not\models \exists F^{ini}.\pi(ec')$, we cannot extend $p$ into a path $p.ec'$ carrying $\sigma.ev$ through $\nu'$, i.e., such that $\nu' \models \exists F^{ini}.\pi(ec')$ and $\nu'(p.ec')=\sigma.ev$. Then $\sigma.ev\not\in Traces(\mathbb{G})$. If $chan(ec')\in C^{out} \cup UC^{in}$, those successors $ec'$ can result in traces ending with quiescence events in $Sem(\mathbb{G})$. Such quiescent traces are of the form $\sigma.(d',\delta!)$ with $d'< d$ and are computed for $p.ec'$ through $\nu'$, i.e., such that $\nu' \models \exists F^{ini}.\pi(ec')$ and $\nu'(p.ec')=\sigma.(d,c!v)$ or $\nu'(p.ec')=\sigma.(d,c?v)$ with $c=chan(ec')$. Yet as $\nu' \not\models \exists F^{ini}.\pi(ec')$, we cannot infer such traces in $Sem(\mathbb{G})$. Then, $\sigma.ev\not\in Sem(\mathbb{G})$.
\qed




\end{document}